\documentclass [12pt,english,a4paper]{article}
\usepackage{amsmath}
\usepackage{amssymb}
\usepackage{slashed}
\usepackage[dvips]{graphics}
\usepackage[dvips]{color}
\usepackage{graphicx}
\newcommand{\be}{\begin{equation}} \newcommand{\ee}{\end{equation}}
\newcommand{\ba}{\begin{eqnarray}} \newcommand{\ea}{\end{eqnarray}}
\newcommand{\bes}{\begin{equation*}} \newcommand{\ees}{\end{equation*}}
\newcommand{\bas}{\begin{eqnarray*}} \newcommand{\eas}{\end{eqnarray*}}
\newcommand{\nn}{\nonumber}
\begin{document}
\centerline{\bf \Large Two Photon Exchange Contributions to Elastic}
\centerline{\bf \Large $ep$ Scattering in the Nonlocal Field Formalism}

\vspace{15mm}

\centerline{\bf  Pankaj Jain, Satish D. Joglekar and Subhadip Mitra}
\centerline{\bf Department of Physics, IIT Kanpur - 208016, India}
\vspace{25mm}

\centerline{\bf Abstract}\bigskip

\noindent We construct a nonlocal gauge invariant Lagrangian to
model the electromagnetic interaction of proton. The Lagrangian
includes  all allowed operators with dimension up to five.
We compute the two photon exchange contribution to elastic
electron-proton scattering using this effective nonlocal Lagrangian.
The one loop calculation in this model includes the standard
box and cross box diagram with the standard on-shell form of the
hadron electromagnetic vertices. Besides this we find an extra contribution
which depends on an unknown constant. We use experimentally extracted
form factors for our calculation. We find that the
correction to the reduced cross section is slightly nonlinear as a function
of the photon longitudinal polarization $\varepsilon$. The non-linearity
seen is within the experimental error bars of the Rosenbluth data.
The final result completely explains the difference between the form factor
ratio $G_E/G_M$ extracted by Rosenbluth separation technique at SLAC and
polarization transfer technique at JLAB.
\pagebreak

\section{Introduction}
\begin{figure}
\begin{center}
\includegraphics[scale=0.7]{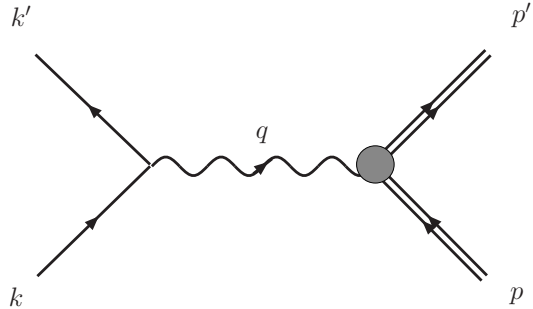}
\end{center}
\caption{\small The one photon exchange diagram contributing to the elastic
electron proton scattering. Here $k$,  $k'$ refer to the initial and final
electron momenta  and $p$, $p'$ to the initial and final proton
momenta respectively. $q = k-k' = p'-p$ is the momentum exchanged.}
\label{fig:diagrams}
\end{figure}

The electromagnetic form factors $F_1$ and $F_2$ parametrize the
vertex of electromagnetic interaction of a photon with
an on-shell proton,
\begin{equation}
\Gamma_{\mu}(p,p')=\gamma_{\mu}F_1(q^{2})+\frac{i\kappa_p
}{2M_p}F_{2}(q^{2})\sigma_{\mu\nu}q^{\nu},\label{eq:OS}
\end{equation}
where $p$ and $p'$ are the initial and final proton momenta,
$M_p$ is the proton mass, $\kappa_p $ its anomalous magnetic moment
and $q= p'-p$ is the momentum transfer. The functions $F_1$ and $F_2$
are called the Dirac and Pauli form factors respectively.
They are experimentally measured
by elastic scattering of electrons on protons, assuming that  the process
is dominated by one photon exchange diagram
(Fig. \ref{fig:diagrams}).
We also define $Q^2 = -q^2\ge 0$.
Besides the form factors $F_1$ and $F_2$, it is also convenient to
define the electric and magnetic form factors (or the Sachs form factors),
$G_E$ and $G_M$ which are more suitable for experimental extraction,
\ba
G_E(Q^2)& =& F_1(Q^2) - \tau\kappa_p F_2 (Q^2)\nonumber\\
G_M(Q^2)& =& F_1(Q^2) + \kappa_p F_2 (Q^2)
\ea
where $\tau = Q^2/4M_p^2$. At $Q^2=0$, $F_1=F_2=1$ and $G_E=G_M/\mu_p=1$, where
$\mu_p$ is the magnetic moment of the proton. The
form factor $G_M\approx \mu_p G_D$ where $G_D$ is the dipole function,
\be G_D = {1\over (1 + {Q^2\over 0.71})^2}. \ee
At low momenta, $G_E$ is also approximately equal to $G_D$.
At large momenta, $Q^2>> 1$ GeV$^2$,
\be G_M, F_1 \propto {1\over Q^4}. \ee
The experimental status of $G_E$ and $F_2$ is, however, currently unclear
at large momentum transfer.

A standard technique for the extraction of the proton form factors
is the Rosenbluth separation \cite{Rosenbluth}. Here one considers
the unpolarized elastic scattering of electrons on target protons.
In the one photon exchange approximation the cross section can be
written as \be {d\sigma\over d\Omega} = {\sigma_{\rm Mott} \over
\varepsilon (1+\tau)} \left[\tau G_M^2(Q^2) + \varepsilon
G_E^2(Q^2)\right] \ee where $\varepsilon=
1/[1+2(1+\tau)\tan^2(\theta_e/2)]$ is the longitudinal
polarization of the photon and $\theta_e$ is the electron
scattering angle. One finds that the reduced cross section,
$\sigma_R = \tau G_M^2(Q^2) + \varepsilon G_E^2(Q^2)$ depends
linearly on $\varepsilon$. By making a linear fit to the observed
$\sigma_R$ as a function $\varepsilon$ at fixed $Q^2$, one can,
therefore, extract both $G_M$ and $G_E$. At large $Q^2$, $G_M$
dominates at all values of $\varepsilon$. Hence the uncertainty in
the extraction of $G_E$ can be large at large $Q^2$. Recent
results for Rosenbluth separation are available from SLAC
\cite{Walker,Andivahis} and JLAB \cite{Qattan}.
The SLAC data shows that ${\mu_p G_E\over G_M}\approx 1$ upto
momentum transfer $Q^2\approx 6$ GeV$^2$. The JLAB data is
available at $Q^2=2.64, 3.20$ and 4.10 GeV$^2$ and shows a similar
trend. This result also implies that the ratio $F_2/F_1\propto
1/Q^2$.

A direct extraction of the ratio $G_E/G_M$ is possible by elastic
scattering of longitudinally polarized electrons on target proton
$\vec e+p\rightarrow e + \vec p$ \cite{Polarization}. In the
one-photon exchange approximation, the recoiling proton acquires
only two polarization components, $P_l$, parallel to the proton
momentum and $P_t$, perpendicular to the proton momentum in the
scattering plane. The ratio,
\be
{G_E\over G_M} = -{P_t\over P_l}{E_e+E_e'\over 2M_p} \tan\left({\theta_e\over 2} \right)
\ee
where $E_e$ and $E_e'$ are the energies of the initial and final
electron. This technique, therefore, directly yields the ratio
$G_E/G_M$. The results \cite{Jones,Gayou,Punjabi}, available from JLAB,
show $\mu_p G_E/G_M$ decreases with $Q^2$. A straight line fit to
the data gives
\be
{\mu_p G_E\over G_M} \approx 1.06 - 0.15\,Q^2
\label{mugegmratio}
\ee in the momentum range $0.5< Q^2< 5.6
$ GeV$^2$. The ratio, therefore, becomes as small as 0.2 at
$Q^2=5.6$ GeV$^2$, the maximum momentum transfer in this
experiment.  The polarization transfer results also imply that
$QF_2/F_1\sim 1$ for $Q^2>1$ GeV$^2$. The observed trend in the
polarization transfer experiment is, therefore, completely
different from what is measured using the Rosenbluth separation.
This is clearly a serious problem and has attracted considerable
attention in the literature \cite{ARZ,PPV}.

\section{Two Photon Exchange}
An obvious source of error is the higher order corrections to the
elastic scattering process. A reliable extraction of the form factors
requires a careful treatment of the radiative corrections including the
soft photon emission, which give a significant correction to the cross
section \cite{Ent,Tsai,Mo,Maximon}.
These contributions are calculated by keeping only the
leading order terms in the soft photon momentum. Furthermore
only the infrared divergent terms,
which are required to cancel the divergences in the soft photon emission,
are included in the radiative corrections.
It is possible that the terms not included in these calculation
may be responsible for the observed difference. Any such correction
is likely to be small and hence cannot significantly change the
results of the polarization transfer experiment. However a small correction
to the Rosenbluth separation could imply a large correction to the
extracted form factor $G_E$. A possible correction is the two photon
exchange diagram which has attracted considerable attention in the literature
\cite{Guichon,Arrington,Tjon,Brodsky,Rekalo}.
Such a diagram is taken into account while computing
the radiative corrections, but only the infrared divergent contribution
is included. It is possible that the remaining contribution gives a
significant correction. One may also consider next to leading
order corrections in the soft photon momenta to the soft photon
emission diagrams. Both of these contributions receive unknown hadronic
corrections and cannot be calculated in a model independent manner.

\begin{figure}
\begin{center}
\begin{tabular}{cc}
\includegraphics[scale=0.7]{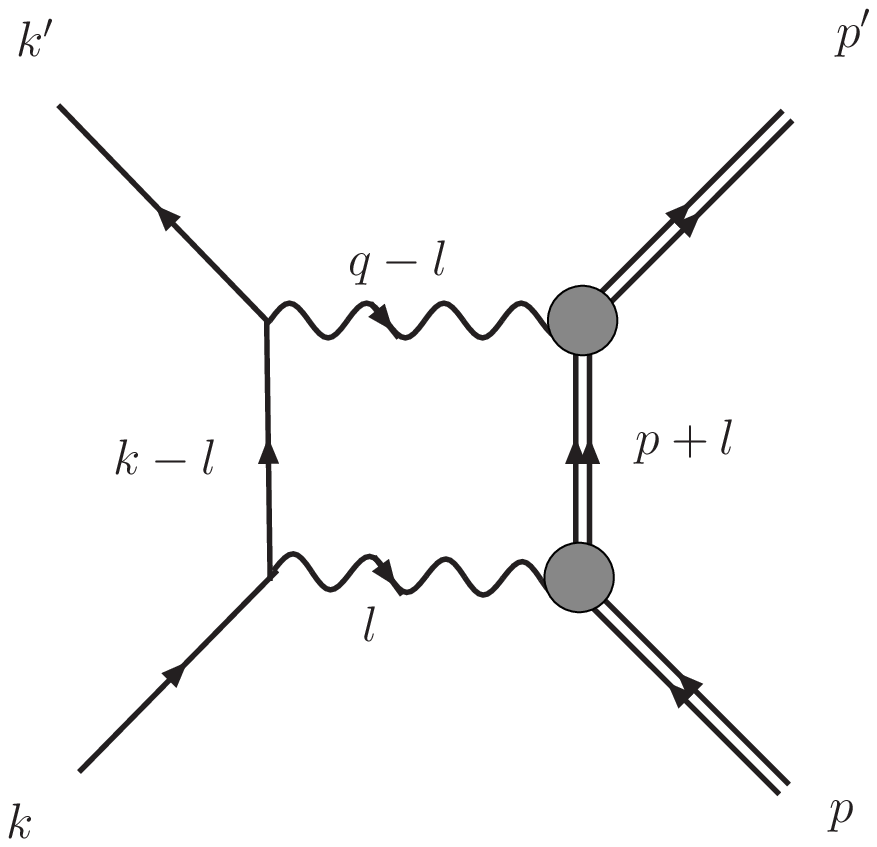}&\includegraphics[scale=0.7]{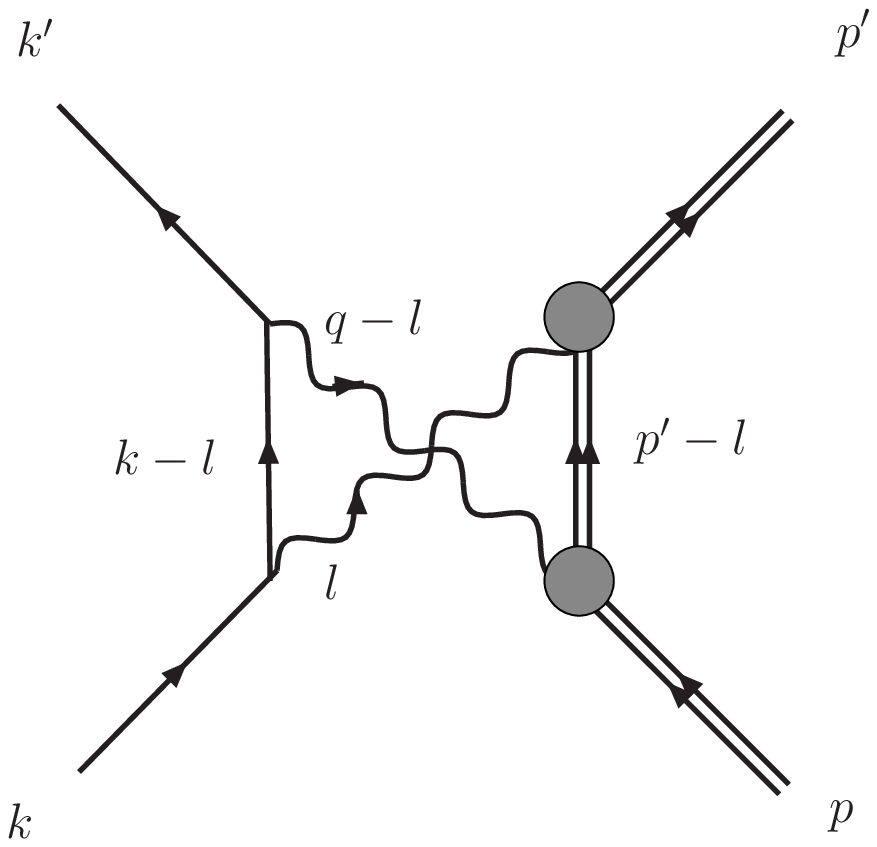}\\
(a)&(b)\\
&
\end{tabular}
\caption{\small The two photon exchange diagrams contributing to the
elastic electron proton scattering: (a) box diagram and (b) cross-box
diagram.}
\label{twophotonbcb}
\end{center}
\end{figure}

In this paper we estimate the two photon exchange contribution
using an effective non-local Lagrangian. The box and cross-box
diagrams which contribute are shown in Fig. \ref{twophotonbcb}(a) and
\ref{twophotonbcb}(b) respectively. As discussed later, in the non-local
formalism we need to evaluate one more diagram. The two photon contribution has
also been obtained by model calculations in Ref.
\cite{Tjon,Brodsky}. The authors find that they are able to
partially reconcile the discrepancy. The results of
Ref. \cite{Tjon,Brodsky} show that the predicted
Rosenbluth plots are no longer linear in $\varepsilon$.
The experimental results obtained from JLAB \cite{Qattan}
show very little deviation from linearity. The SLAC results \cite{Andivahis}
can incorporate some non-linearity due to the presence of relatively
larger errorbars. The present limit on the deviation from linearity is given
in Ref. \cite{Qattan2}. In Ref. \cite{Rekalo} the
authors argue, using charge conjugation and crossing symmetry,
that two photon exchange contribution must necessarily be
nonlinear in $\varepsilon$. If the two photon exchange contribution shows large
non-linearity as a function of $\varepsilon$ then it cannot provide an
an explanation of the observed anomaly.

\section{General electromagnetic vertex of proton}

The elementary electromagnetic vertex of an on-shell  proton is
given in eq. \ref{eq:OS}. When the proton is off-shell, the vertex
is expected to be more general. Further, it must satisfy the
WT identity, following from gauge-invariance, that implies a
relation between $\Gamma_{\mu}(p,p')$ and the inverse proton
propagator, $S_{F}^{-1}(p)$. A \emph{local} theory of interaction
of a proton and a photon would have a \emph{}$U(1)$
gauge-invariance, implied by \emph{local} transformations and
would imply the WT identity:\begin{equation}
q^{\mu}\Gamma_{\mu}(p,p')=S_{F}^{-1}(p')-S_{F}^{-1}(p).\label{eq:WT}\end{equation}
This identity would be violated if in calculating the two photon exchange
diagrams one uses the standard on-shell form factors defined in eq. \ref{eq:OS}
and a free proton propagator.
Here we are interested in formulating the theory in terms of an
effective \emph{nonlocal} action, which will allow us to maintain gauge
invariance in the presence of form factors in the electromagnetic
interaction of proton. It is certainly possible to maintain gauge invariance
in a local theory also but in this case the form factors will arise only
after we take into account loop corrections in strong interactions. It
is not clear how to systematically do calculations in such a case.
In the present case the form factors are present
at the tree level interaction of photon with proton.
The vertex $\Gamma_{\mu}(p,p')$ satisfies a generalized
\emph{non-local} version of the WT identity\footnote{Such non-local
WT identities generally occur in non-local quantum field theories.
See e.g. Ref. \cite{KW92}. This WT identity reduces to the usual one as
$q \rightarrow 0$, provided $g(0) = 1$.}:
\be
g\left(q^{2}\right)q^{\mu}\Gamma_{\mu}(p,p')=S_{F}^{-1}(p')-S_{F}^{-1}(p).\label{eq:NLWT}
\ee
where $g\left(q^{2}\right)$ is a function of $q^{2}$ appearing in
the gauge-transformation equations, ultimately to be related to a
form-factor in the next section. As we shall see in the next
section, this identity follows from a non-local electromagnetic
invariance, and in fact is more appropriate for an extended object
like a proton. In the local limit, the function $g(q^2)\rightarrow 1$, and the
identity in eq. \ref{eq:NLWT} reduces to the local
WT identity. On account of the charge-conjugation invariance of
the proton-photon interaction, the vertex $\Gamma_{\mu}(p,p')$,
a $4\times4$ matrix, must satisfy%
\footnote{The negative signs for momenta on the right-hand-side are
a consequence
of our different sign convention regarding the incoming particle (incoming
momentum positive) and the outgoing particle (out-going momentum
positive).%
}\begin{equation}
C^{-1}\Gamma_{\mu}(p,p')C=-\Gamma_{\mu}^{T}(-p',-p)\label{eq:CC}\end{equation}
where $C$ is the charge-conjugate matrix, with
$C\gamma_{\mu}C^{-1}=-\gamma_{\mu}^{T}$ \cite{BD}. We now express
the vertex in its most general form, employing the 16 linearly
independent Dirac
matrices: $1, \gamma_5, \gamma_{\mu}, \gamma_{\mu}\gamma_5, \sigma_{\mu\nu}$
and the 4-vectors:
$P^{\mu}\equiv\left(p+p'\right)^{\mu}, q^{\mu}\equiv(p'-p)^{\mu}$.
\ba
\Gamma_{\mu}(p,p') & = & aP_{\mu}+bq_{\mu}+c\gamma_{\mu}\nonumber \\
& + & d\slashed{P}P_{\mu}+e\slashed{P}q_{\mu}+f\slashed{q}P_{\mu}+g\slashed{q}q_{\mu}\nonumber \\
& + & h\sigma_{\mu\alpha}P^{\alpha}+j\sigma_{\mu\alpha}q^{\alpha}\nonumber \\
& + & k\sigma_{\alpha\beta}P^{\alpha}q^{\beta}P_{\mu}+l\sigma_{\alpha\beta}P^{\alpha}q^{\beta}q_{\mu}
+m\gamma_{\alpha}\gamma_5\xi_{\mu}^{\nu\alpha\beta}P_{\beta}q_{\nu}.\label{eq:GEN}
\ea
Here, the 12 coefficients $a, b, ...., m$ are functions of the three
Lorentz invariants $p^{2}, p'^{2}, q^{2}$. Charge conjugation
requires that $a, c, d, g, j, k$ are \emph{symmetric} under
$p^{2}\leftrightarrow p'^{2}$ and $b, e, f, h, l, m$ are antisymmetric
under the same operation. To implement the WT identity, we
express,\begin{equation}
S_{F}^{-1}(p)=\alpha\left(p^{2}\right)\slashed{p}+\beta\left(p^{2}\right)\label{eq:IP}\end{equation}
in its most general form. We then impose the WT identity given in
eq. \ref{eq:NLWT}. This leads to some constraints between the
coefficients. The net result of all this is to yield the following
form for $\Gamma_{\mu}$:
\be
\nn \Gamma_{\mu}(p,p') =
a'P_{\mu}+c'\gamma_{\mu}+j\sigma_{\mu\alpha}q^{\alpha}+d'\slashed{P}P_{\mu}
+7 \mbox{ divergence free terms}.
\ee
We enumerate the divergence free (i.e. $X_{\mu}$ with $q^{\mu}X_{\mu}\equiv0$) terms:
\ba
&&b'\left[(p'^{2}-p^{2})q_{\mu}-q^{2}P_{\mu}\right]+f\left(-P.q\gamma_{\mu}+\slashed{q}P_{\mu}\right)+g\left(-q^{2}\gamma_{\mu}+\slashed{q}q_{\mu}\right)\nonumber\\
&+&k\left(-\sigma_{\mu\alpha}P^{\alpha}P.q+\sigma_{\alpha\beta}P^{\alpha}q^{\beta}P_{\mu}\right)+l\left(\sigma_{\mu\alpha}P^{\alpha}q^{2}-\sigma_{\beta\alpha}P^{\alpha}q^{\beta}q_{\mu}\right)\nonumber \\
&+&e'\left(-q^{2}\slashed{P}P_{\mu}+P.q\slashed{P}q_{\mu}\right)+m\gamma_{\alpha}\gamma_5\xi_{\mu}^{\nu\alpha\beta}P_{\beta}q_{\nu}.\nonumber
\ea
We further note the relations that arise from the WT identity
and which restrict the \emph{form} of some of the coefficients
$(a',c')$ considerably:
\begin{eqnarray}
a'=\frac{\beta\left(p'^{2}\right)-\beta\left(p^{2}\right)}{g\left(q^{2}\right)\left(p'^{2}-p^{2}\right)};
\;\;c'=\frac{\alpha\left(p'^{2}\right)+\alpha\left(p^{2}\right)}{2g\left(q^{2}\right)};
\;\;d'=\frac{\alpha\left(p'^{2}\right)-\alpha\left(p^{2}\right)}{g\left(q^{2}\right)\left(p'^{2}-p^{2}\right)}.\nonumber
\end{eqnarray}
whereas the coefficients of $j, b, f, g, k, l, m$ are completely
arbitrary functions of the Lorentz invariants. We make several
observations:

\begin{enumerate}
\item We note first that power counting would associate all operators except
those three with coefficients $a', c', j$ with a local operator of
dimension 6 or higher.
\item We note that the dependence on $q^{2}$ of \emph{both} $a'$ and $c'$
are \emph{identical.} Near mass-shell\footnote{the condition that
$S_F(p)\sim \frac1{\slashed{p}-M_p}$ near mass-shell requires that
$\alpha_0M_p+\beta_0=0;\;\alpha_0+2M_p^2\alpha_1+2M_p\beta_1=1$},
$\alpha\left(p^{2}\right)\sim
\alpha_0+\alpha_1[p^{2}-M_p^2],\,\,\beta\left(p^{2}\right)\sim
\beta_0+\beta_1[p^{2}-M_p^2]$; and thus, \begin{eqnarray}
a'&=&\left\{ \beta_1+\mathcal{O}[p^{2}-M_p^2]\right\}
g^{-1}\left(q^{2}\right);\,\,
\nonumber\\
c'&=&\left\{
\alpha_0+\frac1{2}\alpha_1[p^{2}+p'^{2}-2M_p^2]\right\}
g^{-1}\left(q^{2}\right).\label{eq:ac}\end{eqnarray}

\item The on-shell expression (eq. \ref{eq:OS}) for $\Gamma_{\mu}(p, p')$ takes
operators of dimensions 4 (electric) and 5 (magnetic) into account.
It is then logical that the \emph{only} other operator of dimensions
5 should also be included in the \emph{off-shell} expression for the
$\Gamma_{\mu}(p,p')$. We shall take these three terms into account
in our minimal effective Lagrangian model.
\end{enumerate}

\section{Effective Lagrangian Model}

We represent the interaction of the photon-proton system by
an effective nonlocal Lagrangian model based on the discussion in the last
section. We adopt the following guidelines in the process:

\begin{itemize}
\item The Lagrangian model should incorporate up to dimension 5 operators,
for reasons partly explained in the previous section. The assumption is
that in the effective Lagrangian approach, the higher dimension operators
will contribute much less.
This is borne out in the calculations performed. (See Figure 15 and the
subsequent discussion.)

\item The model should incorporate the results regarding the form of the
coefficients $a',c'$ obtained earlier (see eq. \ref{eq:ac}), thus
at least embody the form-factors on mass-shell. The resulting model
is necessarily non-local.
\item We assume that the model has lowest order derivatives of fermions.
Our assumption about the dimensionality of operators is consistent
with this.
\item We require that this \emph{non-local} model has an equivalent
form of gauge-invariance. Such constructions of \emph{non-local} versions
of local symmetries are known in literature \cite{M90,KW92} and we
shall show explicitly that our model below has a very simple form
of non-local gauge-invariance.
\end{itemize}
A Lagrangian model which satisfies these constraints is given by,\[
\mathcal{L=\overline{\psi}}\left(i\slashed{\partial}-ef'_1\left[\frac{\partial^{2}}{\Lambda^{2}}\right]\slashed{A}-M_p\right)\psi+\frac{a''}{2M_p}\mathcal{\overline{\psi}}\left(\sigma_{\mu\nu}f'_{2}\left[\frac{\partial^{2}}{\Lambda^{2}}\right]F^{\mu\nu}\right)\psi\\
+\frac{b''}{2M_p}\mathcal{\overline{\psi}}\widetilde{D}^{2}\psi\]
where
$i\widetilde{D}=i\partial-ef'_1\left[\frac{\partial^{2}}{\Lambda^{2}}\right]A$
is the \emph{non-local covariant} derivative. We point out that the form
factors, $f'_1$ and $f_2'$ are to be extracted directly from experiments.
We make a number of
observations regarding this effective Lagrangian:

\begin{enumerate}
\item $\mathcal{L}$ is invariant under the \emph{non-local} form of gauge
transformations:\[ \delta
A_{\mu}=-\partial_{\mu}\alpha\left(x\right);\psi(x)\rightarrow
e^{ief'_1\left[\frac{\partial^{2}}{\Lambda^{2}}\right]\alpha\left(x\right)}\psi(x),\overline{\psi}(x)\rightarrow\overline{\psi}(x)e^{-ief'_1\left[\frac{\partial^{2}}{\Lambda^{2}}\right]\alpha\left(x\right)}\]
or equivalently,\[ \delta
A_{\mu}=-\partial_{\mu}f'^{-1}_1\left[\frac{\partial^{2}}{\Lambda^{2}}\right]\beta\left(x\right);\psi(x)\rightarrow
e^{ie\beta\left(x\right)}\psi(x),\overline{\psi}(x)\rightarrow\overline{\psi}(x)e^{-ie\beta\left(x\right)}\]
In the latter form, the gauge transformations are similar to the usual local ones, with the exception that in the
first of these $\partial_\mu \rightarrow \partial_\mu f'^{-1}_1\left(\frac{\partial^2}{\Lambda^2}\right)$. This leads
to the non-local WT identity of eq. \ref{eq:NLWT}; i.e., with a replacement $q_\mu \rightarrow q_\mu f'^{-1}_1\left(\frac{-q^2}{\Lambda^2}\right) \equiv q_\mu g(q^2)$ in eq. \ref{eq:WT}. Under this transformation, $F^{\mu\nu}$ and hence the second term
is gauge-invariant independent of the form of $f'_{2}$. Also, the
non-local gauge-covariant derivative
satisfies: $\widetilde{D}\psi\rightarrow
e^{ie\beta\left(x\right)}\widetilde{D}\psi(x)$.
\item The last term generates a term proportional to $P_{\mu}$ in $\Gamma_{\mu}(p,p')$
with a form factor proportional to $f'_1$, the \emph{same}
\emph{one} appearing in the electric term. This is consistent with
the comment on the form of $a'$ and $c'$ given earlier.
\item The (non-local) gauge-invariance of the last term requires that it
is composed of the (non-local) gauge-covariant derivative: this restricts
the form-factor present in this term as above.
\item The Lagrangian is exactly valid as long
as the proton is on-shell, irrespective of the value of the momentum transfer
$q^2$. In this limit the interaction of the proton with photon is described
in terms of the two form factors and no higher dimensional terms are required.
The higher derivative terms we drop give higher order contributions
in powers of $(P^2-M_p^2)/\Lambda^2$, where $P^2-M_p^2$ is the offshellness
of proton momentum. These higher order terms can be dropped as long as
the dominant contribution to a process is obtained from the kinematic region
where $(P^2-M_p^2)/\Lambda^2<<1$.
\item In the limit $\Lambda\rightarrow \infty$ we reproduce the local
field theory model for a proton with an anomalous magnetic moment.
\end{enumerate}
Before we proceed, a comment on the non-local form of
gauge-invariance is in order. It appears that \emph{a local} form
of gauge-invariance for \emph{extended} particles such as a proton
is inappropriate. Consider the wave-function of an extended
particle centered at $\mathbf{x}$: viz. $\psi(\mathbf{x,}t)$. Let
$\mathbf{y}$ be a point within the charge-radius $R$ of the
proton: $|\mathbf{x-y}|<R$. Let us imagine that a
gauge-transformation on $A_{\mu}$ is carried out (at $t$) around
$\mathbf{y}$ with a very narrow support,
$\rho:\rho<<|\mathbf{x-y}|$. In the model of fundamental
constituents, the quark wave-function should be affected around
$\mathbf{y}$, which in turn should affect the proton wave-function
\emph{even though} the gauge-transformation at $\mathbf{x}$,
depending on $\alpha\left(x\right)$ will be zero. Thus, the proton
wave-function should be affected by a local gauge-transformation
with a support anywhere in its charge radius. The above form of
non-local version of gauge-transformations embodies this idea.
Note that the Fourier transform of
$f'_1\left[\frac{\partial^{2}}{\Lambda^{2}}\right]$ has a support
over a distance $\sim1/\Lambda\sim R$.

It proves convenient to rearrange the Lagrangian as follows%
\footnote{Actually, the constant $M_p$ and the normalization of KE term are also
modified below. However, we shall soon modify the form of the Lagrangian
further, where this proves unnecessary.%
} (Recall the relation
$\slashed{D}^{2}=D^{2}+\frac{e}{2}\sigma_{\mu\nu}F^{\mu\nu}$):
\begin{equation}
\mathcal{L=\overline{\psi}}\left(i\slashed{\widetilde{D}}-M_p\right)\psi+
\frac{\tilde a}{2M_p}\mathcal{\overline{\psi}}\left(\sigma_{\mu\nu}f'_{2}\left[\frac{\partial^{2}}{\Lambda^{2}}\right]F^{\mu\nu}\right)\psi\nonumber \\
+\frac{\bar b}{2M_p}\mathcal{\overline{\psi}}\,\left(i\slashed{\widetilde{D}}-M_p\right)^{2}\psi\label{eq:L}
\end{equation}
Had there been no magnetic term, the last term would have formally
vanished by classical equation of motion. We note that an inclusion
of the last term has now modified the inverse propagator: it has non-vanishing
terms at $e=0$. This, in particular, gives a spurious pole in the
propagator at another value of $\slashed{p}$. This problem can be avoided
if we can write the $\mathcal{L}$ in the following form:
\begin{equation}
\mathcal{L=\overline{\psi}}\left(i\slashed{\widetilde{D}}-M_p\right)exp\left\{
\frac{\bar b}{2M_p}\left(i\slashed{\widetilde{D}}-M_p\right)\right\}
\psi+\frac{\bar a}{2M_p}\mathcal{\overline{\psi}}\left(\sigma_{\mu\nu}f'_{2}\left[\frac{\partial^{2}}{\Lambda^{2}}\right]F^{\mu\nu}\right)\psi\label{eq:ModL}
\end{equation}
which is now understood to have been consistently truncated to a
given order in $\bar b$. We now note that the inverse propagator is:\[
\left(\slashed{p}-M_p\right)exp\left\{
\frac{\bar b}{2M_p}\left(\slashed{p}-M_p\right)\right\}
\]
 and has only one zero at $\slashed{p}-M_p=0$ and the residue of the
propagator at the pole is $1$. (In this form of $\mathcal{L}$,
$\bar a$ is related to the anomalous magnetic moment, $\kappa_p$ by the relation,
$\bar a=e\kappa_p/2$ and $M_p$ is the physical
mass). Since we shall consider, in the two photon exchange
calculation, terms with the last operator of dimension 5 inserted
in \emph{two} photon vertices, we shall do the entire calculation
consistently to $O\left({\bar b}^{2}\right)$ using eq. \ref{eq:ModL}. In
this case, the propagator for the proton is
\[
\frac{i}{\slashed{p}-M_p}\,exp\left\{
-\frac{\bar b}{2M_p}\left(\slashed{p}-M_p\right)\right\}
\approx\frac{i}{\slashed{p}-M_p}-\frac{i\bar b}{2M_p}+\frac{i{\bar b}^{2}}{8M_p^{2}}\left(\slashed{p}-M_p\right)\]

\section{Reduction of the action}

In this section, we shall find an effective way to calculate the matrix
elements involving insertion of the last term in the action. Since
a 2-photon exchange diagram at 1-loop is at most $\mathcal{O}[{\bar b}^{2}]$, we
shall evaluate the effect of this term to $\mathcal{O}[{\bar b}^{2}]$ . What we are
interested in are the \emph{tree}-order matrix elements of two (possibly
virtual) photon emission from an \emph{on-shell} proton\emph{.} The
calculation of these can be simplified considerably \emph{in this
context} with the use of the fermion equations of motion. The result
is simple: of all the terms up to $\mathcal{O}[{\bar b}^{2}]$, viz. $\mathcal{O}[\bar b, \bar b
\bar a, {\bar b}^{2}, {\bar b}^{2} \bar a, {\bar b}^{2} {\bar a}^{2}]$,
only the last term of $\mathcal{O}[{\bar b}^{2} {\bar a}^{2}]$ gives a non-zero result.
While the result can be worked out, it is most effectively dealt with
in the path-integral formulation.

We define,\[ W[J^{\mu},K,\overline{K}]=\int D\phi\exp i\left\{
\int
d^4x[\mathcal{L}+J^{\mu}A_{\mu}+\overline{K}\psi+\overline{\psi}K\right\}
\] where $\mathcal{L}$ is the action of eq. \ref{eq:ModL} and
$D\phi$ denotes generically the measure of the path-integral. We
now perform a field transformation:\begin{equation}
\psi=\exp\left\{
-\frac{\bar b}{2M_p}\left(i\slashed{\widetilde{D}}-M_p\right)\right\}
\psi'\label{eq:FT1}\end{equation} Under this transformation, the
Jacobian is,
\begin{eqnarray}
J&=&\det\exp\left\{
-\frac{\bar b}{2M_p}\left(i\slashed{\widetilde{D}}-M_p\right)\right\}
=\exp tr\left\{
-\frac{\bar b}{2M_p}\left(i\slashed{\widetilde{D}}-M_p\right)\right\} \nonumber\\
& =& \exp tr\left\{ \frac{\bar b}{2}\mathcal{I}\right\} =
\mbox{a constant.}
\end{eqnarray}
and hence can be ignored for the connected Green's functions. This
then yields,
\begin{eqnarray*}
&& \hspace{-10mm}W[J^{\mu},K,\overline{K}] \\
&=& \hspace{-3mm} \int D\phi \exp i\Bigg\{ \int
d^4x\left[\mathcal{\overline{\psi}}\left(i\slashed{\widetilde{D}}-M_p\right)\psi'+\frac{\bar a}{2M_p}\mathcal{\overline{\psi}}\left(\sigma_{\mu\nu}f'_{2}\left[\frac{\partial^{2}}{\Lambda^{2}}\right]F^{\mu\nu}\right)
\nonumber\right.\\
&\times & \left.\exp\left\{
-\frac{\bar b}{2M_p}\left[i\slashed{\widetilde{D}}-M_p\right]\right\}
\psi'\right]\ \Bigg\}\times \exp i\Bigg\{ \int
d^4x\biggl[J^{\mu}A_{\mu}+\overline{\psi}K \\
&+& \left.\overline{K}\exp\left\{ -\frac{\bar b}{2M_p}
\left[i\slashed{\widetilde{D}}-M_p\right]\right\} \psi'\right]\
\Bigg\}\\
\end{eqnarray*}

We note that if we had $\bar a=0$, we would have no left-over term in
the \emph{action.} It is easy to show that $\overline{K}\left[\exp\left\{ -\frac{\bar b}{2M_p}\left[i\slashed{\widetilde{D}}-M_p\right]\right\} -1\right]\psi'$
does not contribute to \emph{tree-level on-shell} proton matrix elements%
\footnote{This is because one does not have a pole in at least one of the external
momenta: it is forbidden either by an explicit factor of $\slashed{p}-M_p$,
or by a vertex.%
}. Thus all tree-level on-shell 2-proton matrix elements involving
the last term in eq. \ref{eq:ModL} are at least of order $\bar a \bar b$. We
now expand the action to $\mathcal{O}[{\bar b}^{2}]$. We find,\[
W[J^{\mu},K,\overline{K}]=\int D\phi\exp i\left\{ \int
d^4x\left[\mathcal{L}'+J^{\mu}A_{\mu}+\overline{K}\psi'+\overline{\psi}K\right]\right\}
+\mathcal{O}[{\bar b}^3]\] with
\begin{eqnarray*}
\mathcal{L}'&=&\left[
\mathcal{\overline{\psi}}+\frac{\bar a}{2M_p}\mathcal{\overline{\psi}}\left(\sigma_{\mu\nu}f'_{2}\left[\frac{\partial^{2}}{\Lambda^{2}}\right]F^{\mu\nu}\right)\left\{
-\frac{\bar b}{2M_p}\right.\right.
+ \left.\left.\frac{{\bar b}^{2}}{8M_p^{2}}\left(i\slashed{\widetilde{D}}-M_p\right)\right\}
\right] \left(i\slashed{\widetilde{D}}-M_p\right)\psi'\\
&+&\frac{\bar a}{2M_p}\mathcal{\overline{\psi}}\left(\sigma_{\mu\nu}f'_{2}\left[\frac{\partial^{2}}{\Lambda^{2}}\right]F^{\mu\nu}\right)\psi'
\end{eqnarray*}
We now perform another field transformation,\begin{equation}
\mathcal{\overline{\psi}}+\frac{\bar a}{2M_p}\mathcal{\overline{\psi}}\left(\sigma_{\mu\nu}f'_{2}\left[\frac{\partial^{2}}{\Lambda^{2}}\right]F^{\mu\nu}\right)\left\{
-\frac{\bar b}{2M_p}+\frac{{\bar b}^{2}}{8M_p^{2}}\left(i\slashed{\widetilde{D}}-M_p\right)\right\}
=\mathcal{\overline{\psi}}'\equiv\mathcal{\overline{\psi}}[1+X]\label{eq:FT2}\end{equation}
We can write,
\begin{eqnarray*}
\overline{\psi} & = & \mathcal{\overline{\psi}}'\left\{ 1+\frac{\bar a}{2M_p}\sigma_{\mu\nu}f'_{2}\left[\frac{\partial^{2}}{\Lambda^{2}}\right]F^{\mu\nu}\left\{ -\frac{\bar b}{2M_p}+\frac{{\bar b}^{2}}{8M_p^{2}}\left(i\slashed{\widetilde{D}}-M_p\right)\right\} \right\} ^{-1}\\
& = &
\mathcal{\overline{\psi}}'+\mathcal{\overline{\psi}}'\left(-\frac{\bar a}{2M_p}\right)\sigma_{\mu\nu}f'_{2}\left[\frac{\partial^{2}}{\Lambda^{2}}\right]F^{\mu\nu}\left\{
-\frac{\bar b}{2M_p}+\frac{{\bar b}^{2}}{8M_p^{2}}\left(i\slashed{\widetilde{D}}-M_p\right)\right\}
+\mathcal{O}[F^{2}]
\end{eqnarray*}
 The $\mathcal{O}[F^{2}]$ will not matter for the present calculation of 2-photon
exchange, as it will give a term having 3-photon fields. The Jacobean
for this transformation,
\begin{eqnarray*}
1/J' & = & \det[1+X]=\det\left[1+X+\frac{X^{2}}{2}-\frac{X^{2}}{2}+\mathcal{O}[{\bar b}^3]\right]\\
 & = & \det\left[e^{X}-\frac{X^{2}}{2}+\mathcal{O}[{\bar b}^{3}]\right]=\det e^{X}\det\left[1-e^{-X}\frac{X^{2}}{2}+\mathcal{O}[{\bar b}^3]\right]\\
 & = & \exp[trX]\det\left[1-e^{-X}\frac{X^{2}}{2}\right]=1-tr\left(e^{-X}\frac{X^{2}}{2}\right)+\mathcal{O}[{\bar b}^3]\\
 & = & 1-tr\frac{X^{2}}{2}+\mathcal{O}[{\bar b}^3]\\
 & = & 1-\frac{{\bar a}^{2}{\bar b}^{2}}{32M_p^4}f'_{2}\left[\frac{\partial^{2}}{\Lambda^{2}}\right]F_{\mu\nu}f'_{2}\left[\frac{\partial^{2}}{\Lambda^{2}}\right]F^{\mu\nu}\times\mbox{(constant)}
\end{eqnarray*}
The last term does not contribute to the emission of two photons from
a proton line in the tree approximation. As a result of the transformation
(eq. \ref{eq:FT2}), the action then becomes,
\begin{eqnarray*}
\mathcal{L}''&=&\mathcal{\overline{\psi}}'\left(i\slashed{\widetilde{D}}-M_p\right)\psi'+\frac{\bar a}{2M_p}\mathcal{\overline{\psi}}'\left(\sigma_{\mu\nu}f'_{2}\left[\frac{\partial^{2}}{\Lambda^{2}}\right]F^{\mu\nu}\right)\psi'\\
&+&\left(\frac{\bar a}{2M_p}\right)^{2}\left(\frac{\bar b}{2M_p}\right)\mathcal{\overline{\psi}}'\left(\sigma_{\mu\nu}f'_{2}\left[\frac{\partial^{2}}{\Lambda^{2}}\right]F^{\mu\nu}\right)^{2}\psi'\\
&-&\left(\frac{\bar a}{2M_p}\right)^{2}\left(\frac{\bar b^{2}}{8M_p^{2}}\right)\mathcal{\overline{\psi}}'\left(\sigma_{\mu\nu}f'_{2}\left[\frac{\partial^{2}}{\Lambda^{2}}\right]F^{\mu\nu}\right)\left(i\slashed{\widetilde{D}}-M_p\right)\left(\sigma_{\mu\nu}f'_{2}\left[\frac{\partial^{2}}{\Lambda^{2}}\right]F^{\mu\nu}\right)\psi'+\mathcal{O}[\bar b^3]\label{eq:FinL}
\end{eqnarray*}
and the source term transforms into
\begin{eqnarray*}
\overline{\psi}'[1+X]^{-1}K&=&\overline{\psi}'\left\{ 1+\left(-\frac{\bar a}{2M_p}\right)\sigma_{\mu\nu}f'_{2}\left[\frac{\partial^{2}}{\Lambda^{2}}\right]F^{\mu\nu}\left\{ -\frac{\bar b}{2M_p}+\frac{\bar b^{2}}{8M_p^{2}}\left(i\slashed{\widetilde{D}}-M_p\right)\right\} \right\} K\\
&+& \overline{\psi}'\left(\frac{\bar a}{2M_p}\right)\,\left(\frac{\bar b^{2}}{8M_p^{2}}\right)\sigma_{\mu\nu}f'_{2}\left[\frac{\partial^{2}}{\Lambda^{2}}\right]F^{\mu\nu}\sigma_{\mu\nu}f'_{2}\left[\frac{\partial^{2}}{\Lambda^{2}}\right]F^{\mu\nu}K\\
\end{eqnarray*}
None of these terms contribute to the tree approximation 2-photon
matrix element for reasons similar as before.

\begin{figure}
\begin{center}
\includegraphics[scale=0.7]{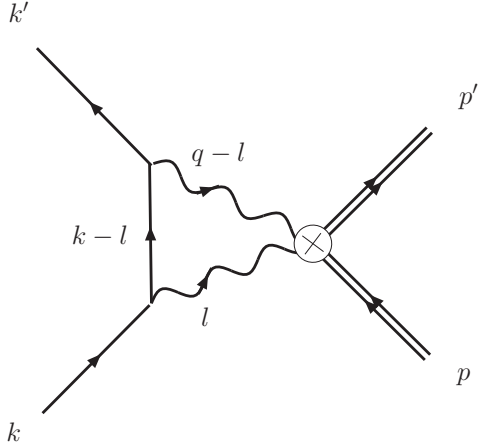}
\caption{\small  The two photon exchange diagram proportional to $\bar b^2$
contributing to the elastic electron proton scattering.
The Feynman rule for this diagram can be obtained from eq. \ref{eqn:bprime}.}
\label{twophotonbpm}
\end{center}
\end{figure}

In conclusion, when we look at the 2-photon exchange diagrams having
up to two insertions of the last term in eq. \ref{eq:L}, each set of
diagrams contain a common part, viz., two (unphysical)-photon tree
amplitude from an on-shell proton. The above discussion shows that
the net effect of that comes from the terms
\begin{eqnarray}
&&\left(\frac{\bar a}{2M_p}\right)^{2}\left(\frac{\bar b}{2M_p}\right)\mathcal{\overline{\psi}}'\left(\sigma_{\mu\nu}f'_{2}\left[\frac{\partial^{2}}{\Lambda^{2}}\right]F^{\mu\nu}\right)^{2}\psi'\nonumber\\
&-&\left(\frac{\bar a}{2M_p}\right)^{2}\left(\frac{\bar b^{2}}{8M_p^{2}}\right)
\mathcal{\overline{\psi}}'
\left(\sigma_{\mu\nu}f'_{2}\left[\frac{\partial^{2}}{\Lambda^{2}}\right]F^{\mu\nu}\right)
\left(i\slashed{\widetilde{D}}-M_p\right)
\left(\sigma_{\mu\nu}f'_{2}\left[\frac{\partial^{2}}{\Lambda^{2}}\right]F^{\mu\nu}\right)
\psi' \label{eqn:bprime}
\end{eqnarray}
We shall show in appendix 1 that the first term does not contribute
in the Feynman gauge. That leaves us with only the last term. The Feynman
diagram corresponding to this term is shown in Fig. \ref{twophotonbpm}.

\section{Calculation and Results}

\begin{figure}
\hskip 1.5cm
\includegraphics[scale=0.5,angle=270]{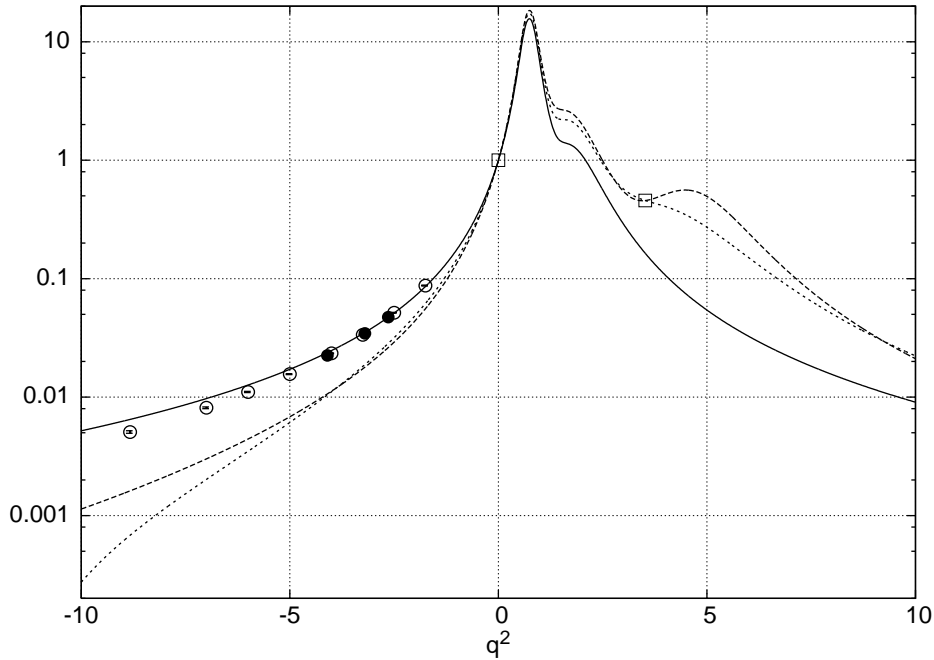}
\caption{\small The amplitude of $G_M/\mu_p$ (solid line) and $G_E$ (dotted line - Model I, dashed line - Model II). The unfilled squares represent the constrained values for $G_E$ at $q^2=0$ and $q^2=4M^2$.
Results of Rosenbluth extraction experiments (filled circles - JLAB, unfilled circles - SLAC) are also shown.}
\label{fig:gmgeamp}
\end{figure}

\begin{figure}
\hskip 1.5cm
\includegraphics[scale=0.5,angle=270]{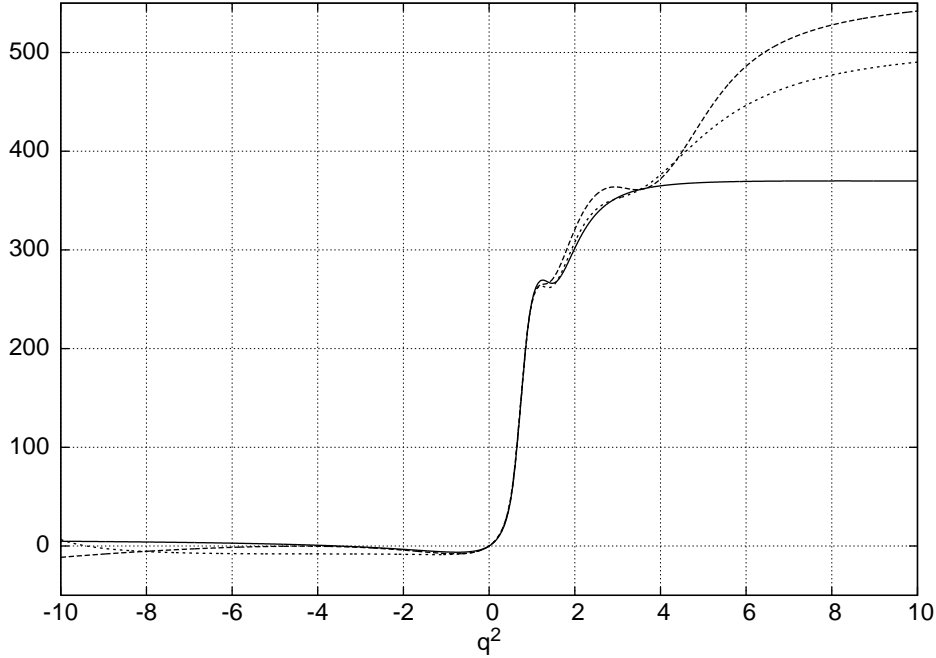}
\caption{\small The phase (in degrees) of $G_M$ (solid line) and $G_E$ (dotted line - Model I, dashed line - Model II).}
\label{fig:gmgmephase}
\end{figure}

In this section we give details of the calculation of the two photon exchange
diagrams using our effective Lagrangian. The calculation
turns out to be complicated due to the explicit presence of
form factors at the vertices. We also require models for the form
factors both in the space-like and time-like regions. In the space
like region the form factor $F_1(q^2)$ is known reasonably well.
In the time like region experimental data exists for the form
factor $G_M(q^2)$ for $4M_p^2< q^2<14$ GeV$^2$, where $4M_p^2$ is
the threshold energy for $p\bar p$ production. In Ref.
\cite{Baldini}, $G_M(q^2)$ has been extracted in the unphysical
region $0< q^2<4M_p^2$ by using dispersion relations
\cite{Mergell,Belushkin}. The
extracted form factor shows two resonances at masses $M\sim 770$
MeV and $M\sim 1600$ MeV. The phase of the magnetic form factor
also shows a large variation in the unphysical region. The
electric form factor $G_E(q^2)$, however, is not well known. The amplitude
in the unphysical region is obtained in Ref. \cite{Baldini_ge}.
However the phase is not known. Our model for the form factors is given in Appendix 2. We use two different models.
Both consist of a sum of simple poles. The corresponding masses and widths
are given in Tables \ref{tab1} and \ref{tab2}. The values of these parameters are obtained
by fits to the experimental data, or the data obtained from experiments
by using dispersion relations \cite{Baldini}.
The resulting amplitude and phase of the form factors for the two models are
shown in Fig. \ref{fig:gmgeamp} and \ref{fig:gmgmephase} respectively.

\begin{figure}
\hskip 2cm
\includegraphics[scale=0.45,angle=270]{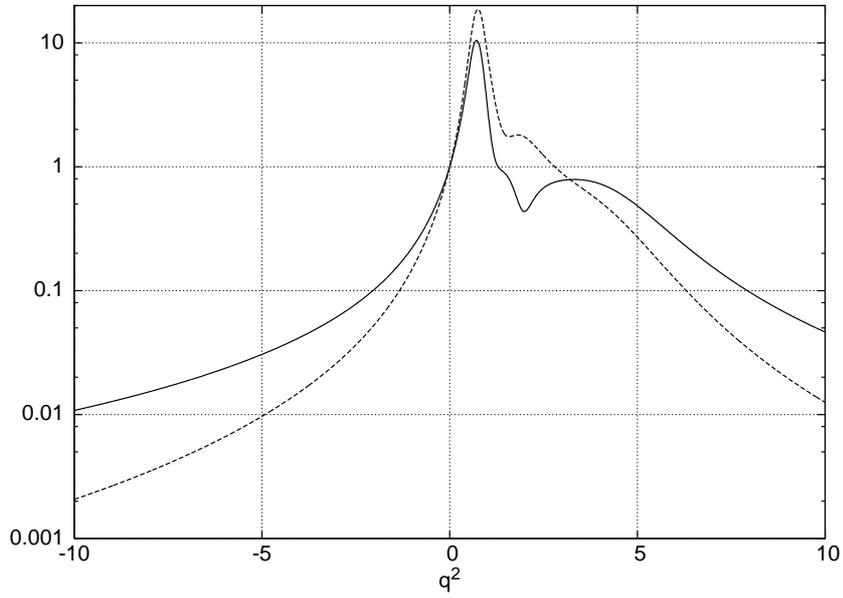}
\caption{\small The amplitude of $F_1$ (solid line) and $F_2$ (dashed line) - Model I.}
\label{fig:f1f2amp1}
\end{figure}

\begin{figure}
\hskip 2cm
\includegraphics[scale=0.45,angle=270]{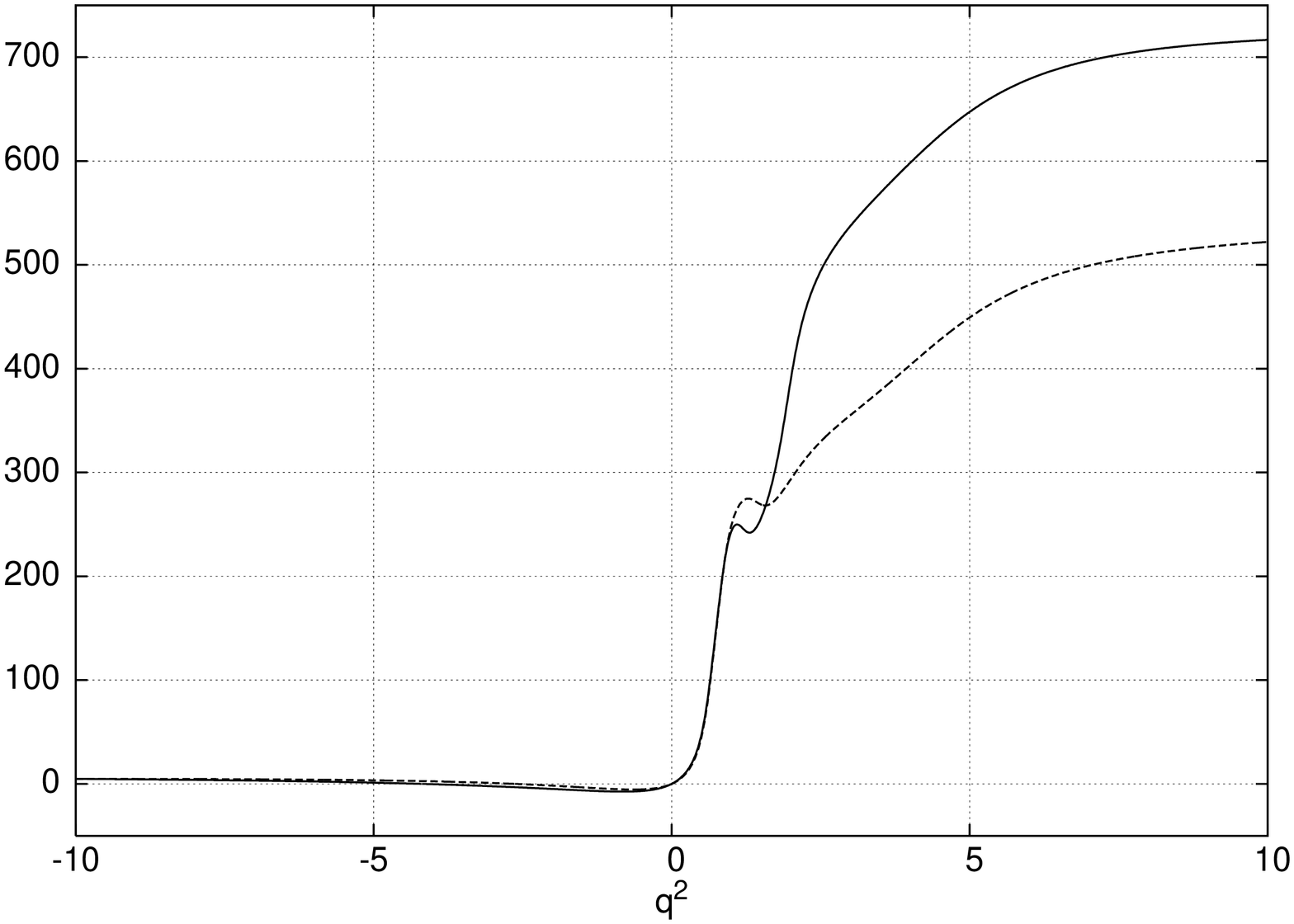}
\caption{\small The phase (in degrees) of $F_1$ (solid line) and $F_2$ (dashed line) - Model I.}
\label{fig:f1f2phase1}
\end{figure}

\begin{figure}
\hskip 2cm
\includegraphics[scale=0.450,angle=270]{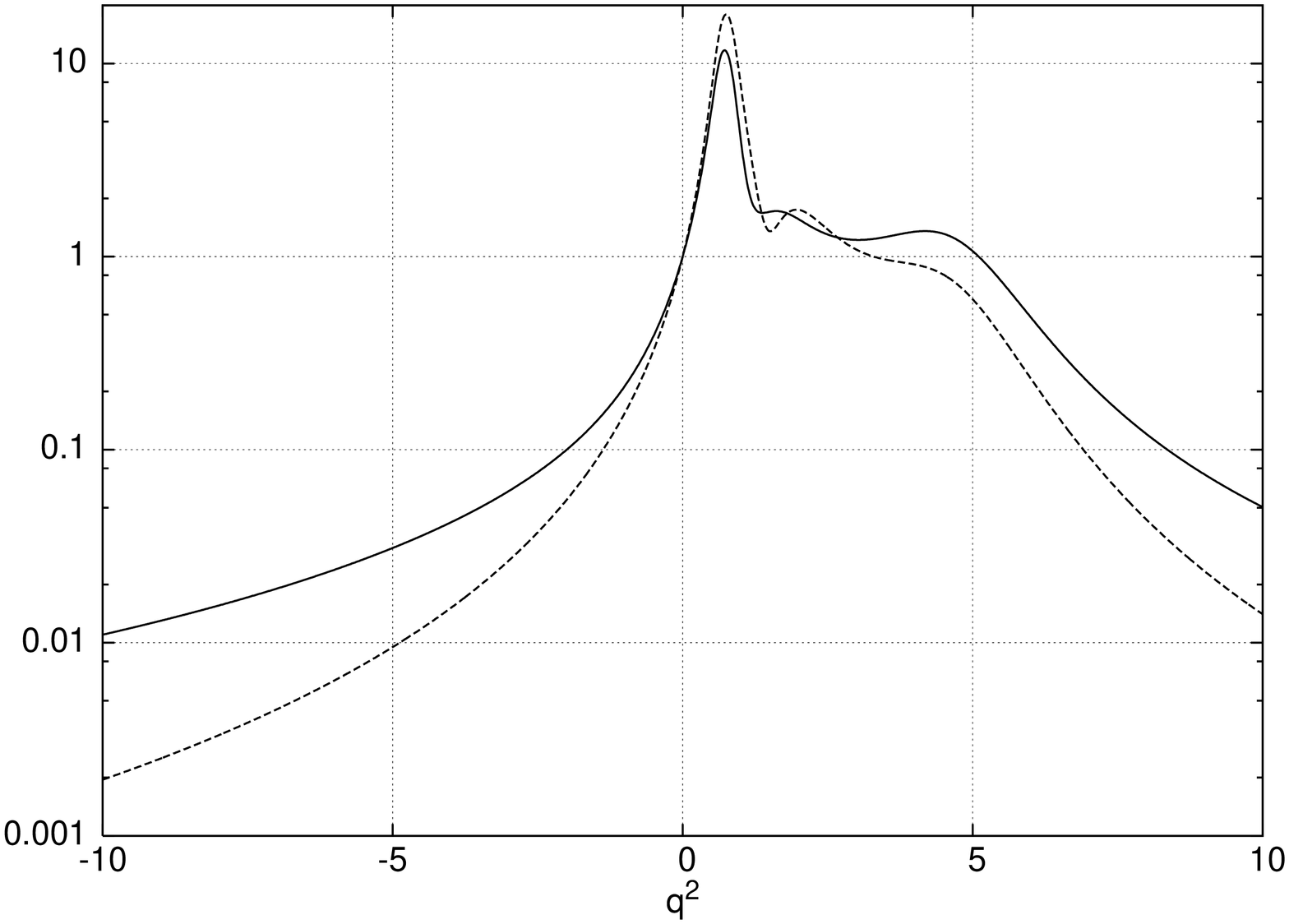}
\caption{\small The amplitude of $F_1$ (solid line) and $F_2$ (dashed line) - Model II.}
\label{fig:f1f2amp2}
\end{figure}

\begin{figure}
\hskip 2cm
\includegraphics[scale=0.450,angle=270]{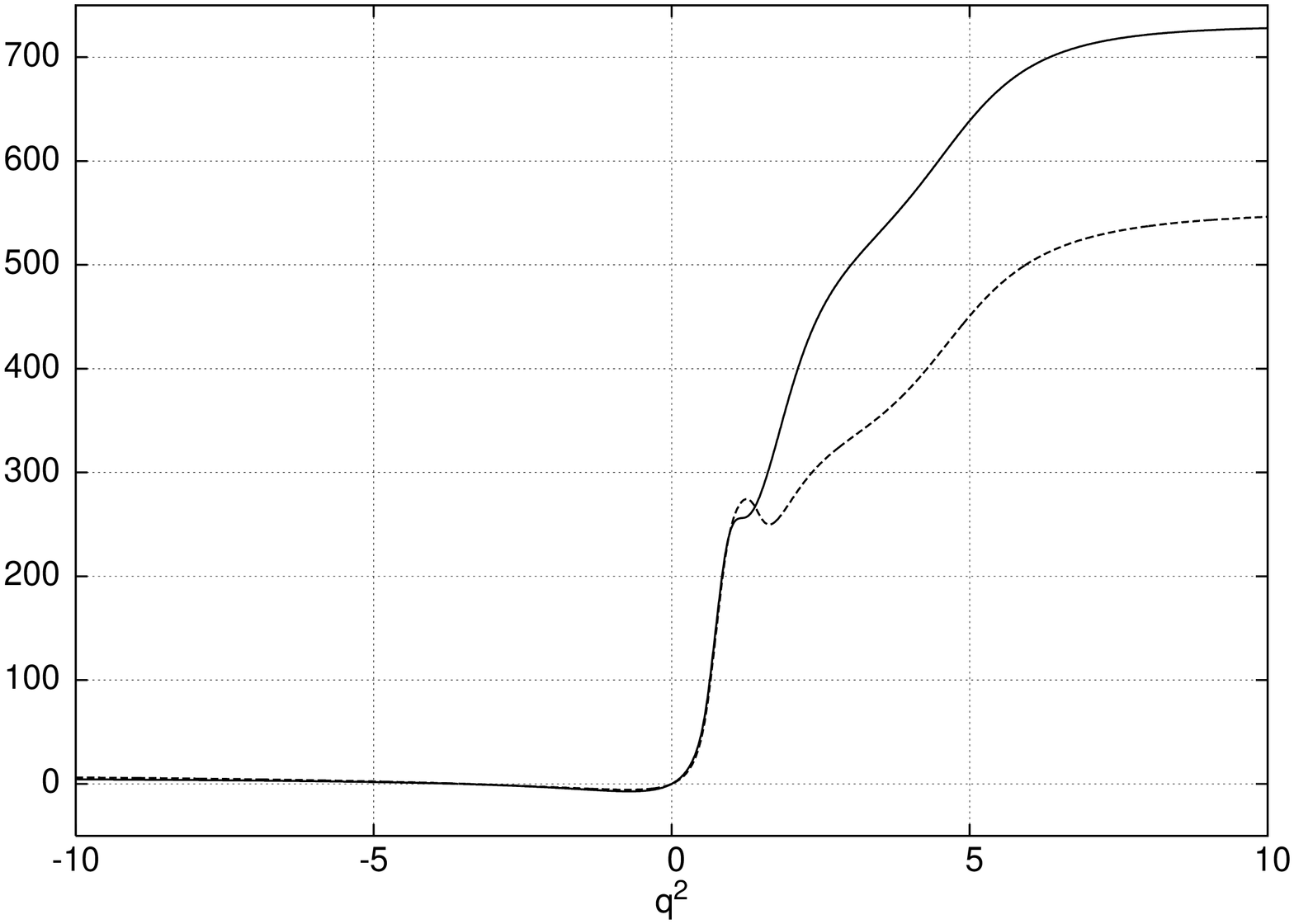}
\caption{\small The phase (in degrees) of $F_1$ (solid line) and $F_2$ (dashed line) - Model II.}
\label{fig:f1f2phase2}
\end{figure}

Using these models for the magnetic and
electric form factors we can obtain the form factors $F_1$ and $F_2$,
required for our calculation.
The models used are convenient since they allow us to use the Feynman
parametrization to compute the loop integrals. The form factors for the two models have a
small imaginary part even for space like momenta. However this
region contributes negligibly to the loop integrals. The dominant
contribution comes from the unphysical region $0<q^2<4M_p^2$
where the form factor is several orders of magnitude larger than its
value in the space like region. In this region
our model provides a very good fit to the extracted form factor
\cite{Baldini}. Moreover the imaginary part in space like region is very
small and unlikely to affect our results significantly. The resulting
amplitudes and phases for model I are shown in Figs. \ref{fig:f1f2amp1}
and \ref{fig:f1f2phase1} respectively. The corresponding results for
model II are shown in Figs. \ref{fig:f1f2amp2}
and \ref{fig:f1f2phase2}.


Using the form factor models in Appendix 2 we can determine the amplitudes of
the box and cross box diagram as well as the amplitude due to the extra term
displayed in eq. \ref{eqn:bprime}. The box diagram amplitude can be written
as,
\ba
\nn i\mathcal{M}_B &=& e^4\sum_{a,b}\int\frac{d^4l}{(2\pi)^4}\frac{\bar{u}(k')\gamma^\mu(\slashed{k}-\slashed{l})\gamma^\nu u(k) }{((k-l)^2 -m_e^2+i\xi)(l^2-\mu^2+i\xi)(\tilde{q}^2-\mu^2 +i\xi)}\\
\nn&\times& \left[\bar{U}(p')\{F_1(\tilde{q})\gamma_\mu + i\frac{\kappa_p}{2M_p}F_2(\tilde{q})\sigma_{\mu\alpha}(\tilde{q})^\alpha\}\right. \frac{\slashed{p}+\slashed{l}+M_p}{(p+l)^2-M_p^2+i\xi}\\
&\times& \left. \{F_1(l)\gamma_\nu + i\frac{\kappa_p}{2M_p}F_2(l)\sigma_{\nu\beta}l^\beta\}U(p)\right],
\ea
where $\tilde{q} = q-l$, $m_e$ is the mass of the electron and
$\xi$ is an infinitesimal positive parameter\footnote{Here we use the
notation $\xi$
instead of the standard notation $\epsilon$ to avoid confusion with the symbol
$\varepsilon$ used to denote the photon longitudinal polarization}.
A small mass of the photon $\mu$ has been introduced in order to
regulate the infrared divergence in these integrals.
The infrared divergent part has to be
subtracted from our result since it is included in the standard radiative
corrections which are applied while extracting the form factor.
Using the form factor model in Appendix 2, we find,
\ba
\nn i\mathcal{M}_B &=& e^4\sum_{a,b}\int\frac{d^4l}{(2\pi)^4}\frac{\bar{u}(k')\gamma^\mu(\slashed{k}-\slashed{l})\gamma^\nu u(k) }{((k-l)^2 -m_e^2+i\xi)(l^2-\mu^2+i\xi)(\tilde{q}^2-\mu^2 +i\xi)}\\
\nn &\times& \left[\bar{U}(p')\{(C_ag_a(\tilde{q})-4M_p^2D_a\tilde{g_a}(\tilde{q}))\gamma_\mu + i2M_pD_a\tilde{g_a}(\tilde{q})\sigma_{\mu\alpha}(\tilde{q})^\alpha\}\right. \frac{\slashed{p}+\slashed{l}+M_p}{(p+l)^2-M_p^2+i\xi}\\
&\times& \left. \{(C_bg_b(l)-4M_p^2D_b\tilde{g_b}(l))\gamma_\nu + i2M_pD_b\tilde{g_b}(l)\sigma_{\nu\beta}l^\beta\}U(p)\right].
\ea
It is convenient to rewrite this expression in terms of the coefficients
$C'$ and $D'$, defined in Appendix 2. We find,
\ba
\nn i\mathcal{M}_B &=& e^4\sum_{i,j}\int\frac{d^4l}{(2\pi)^4}\left[\frac{\bar{u}(k')\gamma^\mu(\slashed{k}-\slashed{l})\gamma^\nu u(k) }{(k-l)^2 -m_e^2+i\xi}\right]\\
\nn & &\times \left[\bar{U}(p')\{(C'_ig_i(\tilde{q})-4M_p^2D'_ig_i(\tilde{q}))\gamma_\mu + i2M_pD'_ig_i(\tilde{q})\sigma_{\mu\alpha}(\tilde{q})^\alpha\}\right.\frac{\slashed{p}+\slashed{l}+M_p}{(p+l)^2-M_p^2+i\xi}\\
& &\times \left. \{(C'_jg_j(l)-4M_p^2D'_jg_j(l))\gamma_\nu + i2M_pD'_jg_j(l)\sigma_{\nu\beta}l^\beta\}U(p)\right]\\
\nn  &\equiv& e^4\int\frac{d^4l}{(2\pi)^4}\frac{1}{((k-l)^2 -m_e^2+i\xi)((p+l)^2-
M_p^2+i\xi)}\\
 & &\times\sum_{i,j} \left[n_1(C'_i,D'_i,C'_j,D'_j) + l^4n_2(C'_i,D'_i,C'_j,D'_j)\right]
\ea
where the last step defines the factors $n_1(C'_i,D'_i,C'_j,D'_j)$ and
$n_2(C'_i,D'_i,C'_j,D'_j)$. We may cancel the $l^2$ factor multiplying
$n_2(C'_i,D'_i,C'_j,D'_j)$ with a factor ($l^2-\mu^2+i\xi$) in the
denominator. We then find,
\ba
\nn i\mathcal{M}_B  &=& e^4\int\frac{d^4l}{(2\pi)^4}\frac{1}{((k-l)^2 -m_e^2+i\xi)((p+l)^2-M_p^2+i\xi)}\\
 & &\times\sum_{i,j} \left[n_1(C'_iD'_i,C'_j,D'_j) + l^2n_2(C'_i,D'_i,C''_j,D''_j)\right].
\ea
The cross-box diagram amplitude can be written as,
\ba
\nn i\mathcal{M}_{CB} &=& e^4\sum_{a,b}\int\frac{d^4l}{(2\pi)^4}\left[\frac{\bar{u}(k')\gamma^\mu(\slashed{k}-\slashed{l})\gamma^\nu u(k) }{(k-l)^2 -m_e^2+i\xi}\right]\left[\frac{1}{(l^2-\mu^2+i\xi)(\tilde{q}^2-\mu^2 +i\xi)}\right]\\
\nn &\times & \left[\bar{U}(p')\{F_1(l)\gamma_\nu + i\frac{\kappa_p}{2M_p}F_2(l)\sigma_{\nu\beta}l^\beta\}\right.\frac{\slashed{p}+\slashed{q}-\slashed{l}+M_p}{(p+\tilde{q})^2-M_p^2+i\xi}\\
\nn &\times&  \left. \{F_1(\tilde{q})\gamma_\mu + i\frac{\kappa_p}{2M_p}F_2(\tilde{q})\sigma_{\mu\alpha}(\tilde{q})^\alpha\}U(p)\right]\\
\nn &=& e^4\sum_{a,b}\int\frac{d^4l}{(2\pi)^4}\left[\frac{\bar{u}(k')\gamma^\mu(\slashed{k}-\slashed{l})\gamma^\nu u(k) }{(k-l)^2 -m_e^2+i\xi}\right]\left[\frac{1}{(l^2-\mu^2+i\xi)(\tilde{q}^2-\mu^2 +i\xi)}\right]\\
\nn &\times&  \left[\bar{U}(p')\{(C_bg_b(l)-4M_p^2D_b\tilde{g_b}(l))\gamma_\nu + i2M_pD_b\tilde{g_b}(l)\sigma_{\nu\beta}l^\beta\}\right.\\
\nn &\times&  \frac{\slashed{p}+\slashed{q}-\slashed{l}+M_p}{(p+\tilde{q})^2-M_p^2+i\xi}\\
 &\times&  \left. \{(C_ag_a(\tilde{q})-4M_p^2D_a\tilde{g_a}(\tilde{q}))\gamma_\mu + i2M_pD_a\tilde{g_a}(\tilde{q})\sigma_{\mu\alpha}(\tilde{q})^\alpha\}U(p)\right].
\ea
The amplitude proportional to ${\bar b}^{2}$ is given by:
\ba
\nn i\mathcal{M}_{\bar b} &=& \left(\frac{e^4\bar b^{2}}{8M_p^2}\right)\sum_{a,b}\int\frac{d^4l}{(2\pi)^4}\left[\frac{\bar{u}(k')\gamma^\mu(\slashed{k}-\slashed{l})\gamma^\nu u(k) }{(k-l)^2 -m_e^2+i\xi}\right]\left[\frac{1}{(l^2-\mu^2+i\xi)(\tilde{q}^2-\mu^2 +i\xi)}\right]\\
\nn &\times& \bar{U}(p')\left[ \left(\frac{i\kappa_p}{2M_p}F_2(\tilde{q})\sigma_{\mu\alpha}(\tilde{q})^\alpha\right)
(\slashed{p}+\slashed{l}-M_p) \left(\frac{i\kappa_p}{2M_p}F_2(l)\sigma_{\nu\beta}l^\beta\right)\right.\\
\nn &+&\left. \left(\frac{i\kappa_p}{2M_p}F_2(l)\sigma_{\nu\beta}l^\beta\right)
(\slashed{p}+\slashed{q}-\slashed{l}-M_p) \left(\frac{i\kappa_p}{2M_p}F_2(\tilde{q})\sigma_{\mu\alpha}(\tilde{q})^\alpha\right)\right]U(p)\\
\nn &=& \left(\frac{e^4\bar b^{2}}{8M_p^2}\right)\sum_{i,j}\int\frac{d^4l}{(2\pi)^4}\left[\frac{\bar{u}(k')\gamma^\mu(\slashed{k}-\slashed{l})\gamma^\nu u(k) }{(k-l)^2 -m_e^2+i\xi}\right]\\
\nn &\times& \bar{U}(p')\left[(i2M_pD'_ig_i(\tilde{q})\sigma_{\mu\alpha}(\tilde{q})^\alpha)(\slashed{p}+\slashed{l}-M_p)
(i2M_pD'_jg_j(l)\sigma_{\nu\beta}l^\beta)\right.\\
& +& \left. (i2M_pD'_jg_j(l)\sigma_{\nu\beta}l^\beta)(\slashed{p}+\slashed{q}-\slashed{l}-M_p)
(i2M_pD'_ig_i(\tilde{q})\sigma_{\mu\alpha}(\tilde{q})^\alpha)\right]U(p).
\ea
In our numerical calculation we set the mass of the electron $m_e=0$.

The contribution of the two photon exchange diagrams to the electron-proton
elastic scattering cross section can be written as
\be
\frac{d\sigma^{2\gamma}}{d\Omega_e} = \frac{2Re(\overline{\mathcal{M}^*_0\mathcal{M}_{2\gamma}})E_e^{'2}}
{64M_p^2\pi^2E_e^2}+\mathcal{O}(\alpha^4).
\ee
where,
\be
\mathcal{M}_{2\gamma} = \mathcal{M}_{B} + \mathcal{M}_{CB} + \mathcal{M}_{\bar b},
\ee
is the total amplitude of the two photon exchange
diagrams and
\be
\mathcal{M}_0 = -\frac{e^2}{q^2}\bar{u}(k')\gamma^\mu
u(k)\bar{U}(p')\left(F_1(q)\gamma_\mu + \frac{i\kappa_p}{2M_p}
F_2(q)\sigma_{\mu\alpha}q^\alpha\right)U(p)
\ee
the tree amplitude. Hence, the contribution of the two photon exchange diagrams to the reduced
cross section is given by:
\ba
\nn \sigma_R^{2\gamma} &=& \frac{\varepsilon(1+\tau)}{\sigma_{Mott}}\frac{d\sigma^{2\gamma}}{d\Omega_e}\\
\nn &=&\left(\frac{4E_e^3sin^4\frac{\theta_e}{2}}{\alpha^2E'_ecos^2\frac{\theta_e}{2}}\right)\varepsilon(1+\tau)
\left(\frac{2Re(\overline{\mathcal{M}^*_0\mathcal{M}_{2\gamma}})E_e^{'2}}{64M_p^2\pi^2E_e^2}\right)\\
&=&\frac{q^4\varepsilon(1+\tau)}{32\alpha^2\pi^2M_p^2(q^2+4E_eE'_e)}
Re(\overline{\mathcal{M}^*_0\mathcal{M}_{2\gamma}}).
\ea

The diagram proportional $\bar b^2$ has no infrared (IR) divergent term. So the contribution coming
from it is computed keeping $\mu^2 = 0$. Contribution from box and cross-box diagrams are computed
at 10 different values of $\mu^2$ (from 0.005 to 0.0095). The numerical calculation of the cross-box
diagram is straightforward since the integral is well defined. However for the
evaluation of the box diagram the numerical evaluation is facilitated
by keeping a small imaginary term $\xi$ in the propagators.
This makes the integral in the infrared limit well defined in the case $m_e=0$.
For each value of $q^2$,
$\varepsilon$ and $\mu^2$ we have calculated
the box diagram amplitude for 4 different values of $\xi$
(between 0.001 and 0.00175).
The amplitudes depends almost linearly with $\xi$. The final $\mu^2$ dependent box diagram amplitudes
are obtained by extrapolation to $\xi=0$. The two different models for the form factors described in
Appendix 2 gives almost identical results. So for the rest of the section we quote the result
obtained using Model-I only.

The IR behaviour of the two photon diagrams have been calculated by Mo and Tsai \cite{Tsai,Mo}.
In the limit $\mu^2 \rightarrow 0$ the leading term from the box and cross-box diagram can be expressed as:
\ba
\mathcal{M}^{2\gamma}_{IR} = \frac{\alpha}{\pi}[K(p',k)-K(p,k)]\mathcal{M}_0,
\ea
where
\bas
K(p_i,p_j) = (p_i.p_j)\int_0^1\frac{dx}{(xp_i+(1-x)p_j)^2}\,
\ln\left[\frac{(xp_i+(1-x)p_j)^2}{\mu^2}\right].
\eas
The IR contribution to the reduced cross section coming from the box and cross-box diagram is given by:
\ba
\sigma^{2\gamma}_{IR} = \frac{2\alpha}{\pi}[K(p',k)-K(p,k)]\sigma^{1\gamma}_R.
\ea
Let
\ba
\sigma^{2\gamma}_{IR}\equiv a_{ir} + b_{ir}\ln\mu^2.
\ea

To remove the IR part from $\sigma^{2\gamma}_R$ we fit it with the following function:
\ba
\sigma^{2\gamma}_R = a(\mu^2) + b(\mu^2)\ln\mu^2
\ea
with
\bas
a(\mu^2) &=& a_0 + a_{ir} + a_1\mu^2 + \mathcal{O}[\mu^4]\\
b(\mu^2) &=& b_{ir} + b_1\mu^2 + \mathcal{O}[\mu^4].
\eas
Here $a_0$ gives the IR removed
$\sigma_R^{2\gamma}$. It has been explicitly verified that keeping
$\mathcal{O}[\mu^4]$ terms in $a(\mu)$ and $b(\mu)$ has no effect on the slope (thus on $G_E$)
of $\sigma_R^{2\gamma}(\varepsilon)$ with respect to $\varepsilon$. The difference
between these two fits leads to a very small correction to $G_M$ only and hence can be ignored.

\begin{figure}
\hskip 1.5cm
\includegraphics[scale=0.75,angle=0]{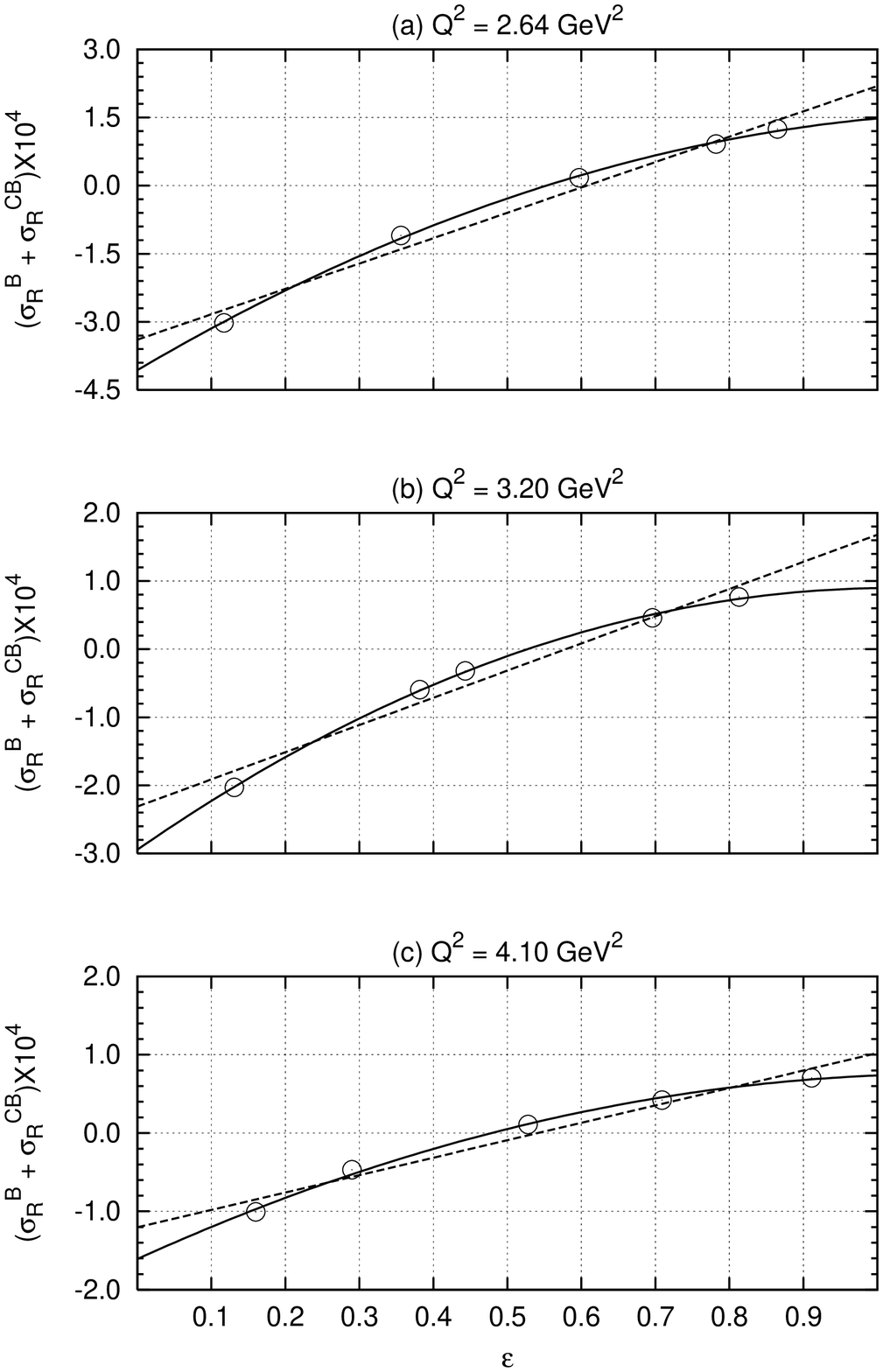}
\caption{\small Total contribution of the box and cross-box diagram to the elastic electron
proton scattering for $Q^2$ = 2.64 GeV$^2$ (a),
3.20 GeV$^2$ (b), 4.10 GeV$^2$ (c). The dashed lines represent $f_1$ (eqn. \ref{eq:result.fit.1})
and the solid curves represent $f_2$ (eqn. \ref{eq:result.fit.2}). The fitting parameters are given
in Table \ref{tab:bcbfit}.}
\label{fig:bcb}
\end{figure}

\begin{figure}
\hskip 1.5cm
\includegraphics[scale=0.75,angle=0]{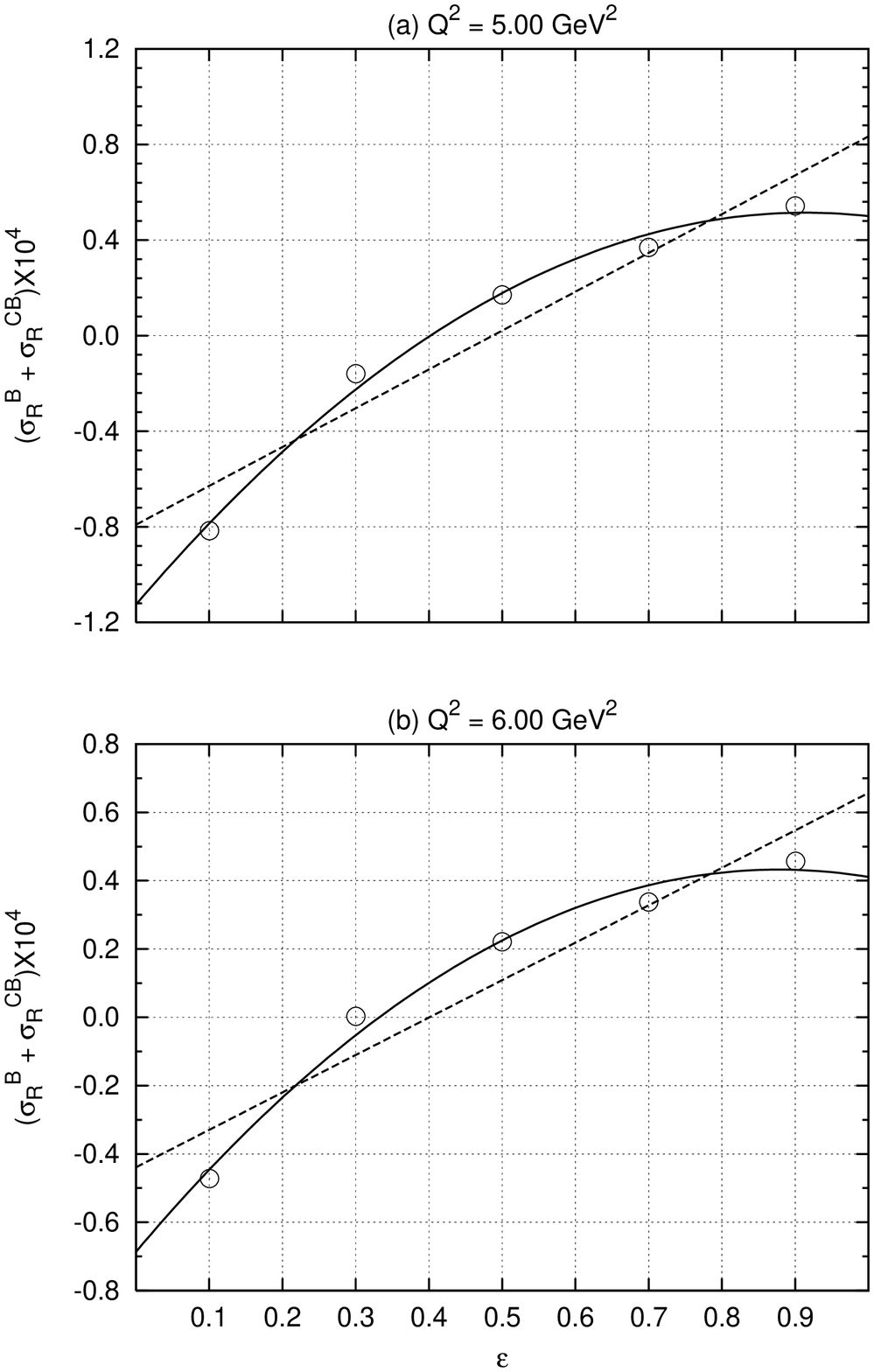}
\caption{\small Total contribution of the box and cross-box diagram to the elastic electron
proton scattering for $Q^2$ = 5 GeV$^2$ (a),
6 GeV$^2$ (b). The dashed lines represent $f_1$ (eqn. \ref{eq:result.fit.1})
and the solid curves represent $f_2$ (eqn. \ref{eq:result.fit.2}). The fitting parameters are given
in Table \ref{tab:bcbfit}.}
\label{fig:bcb.2}
\end{figure}

\begin{figure}
\hskip 1.5cm
\includegraphics[scale=0.75,angle=0]{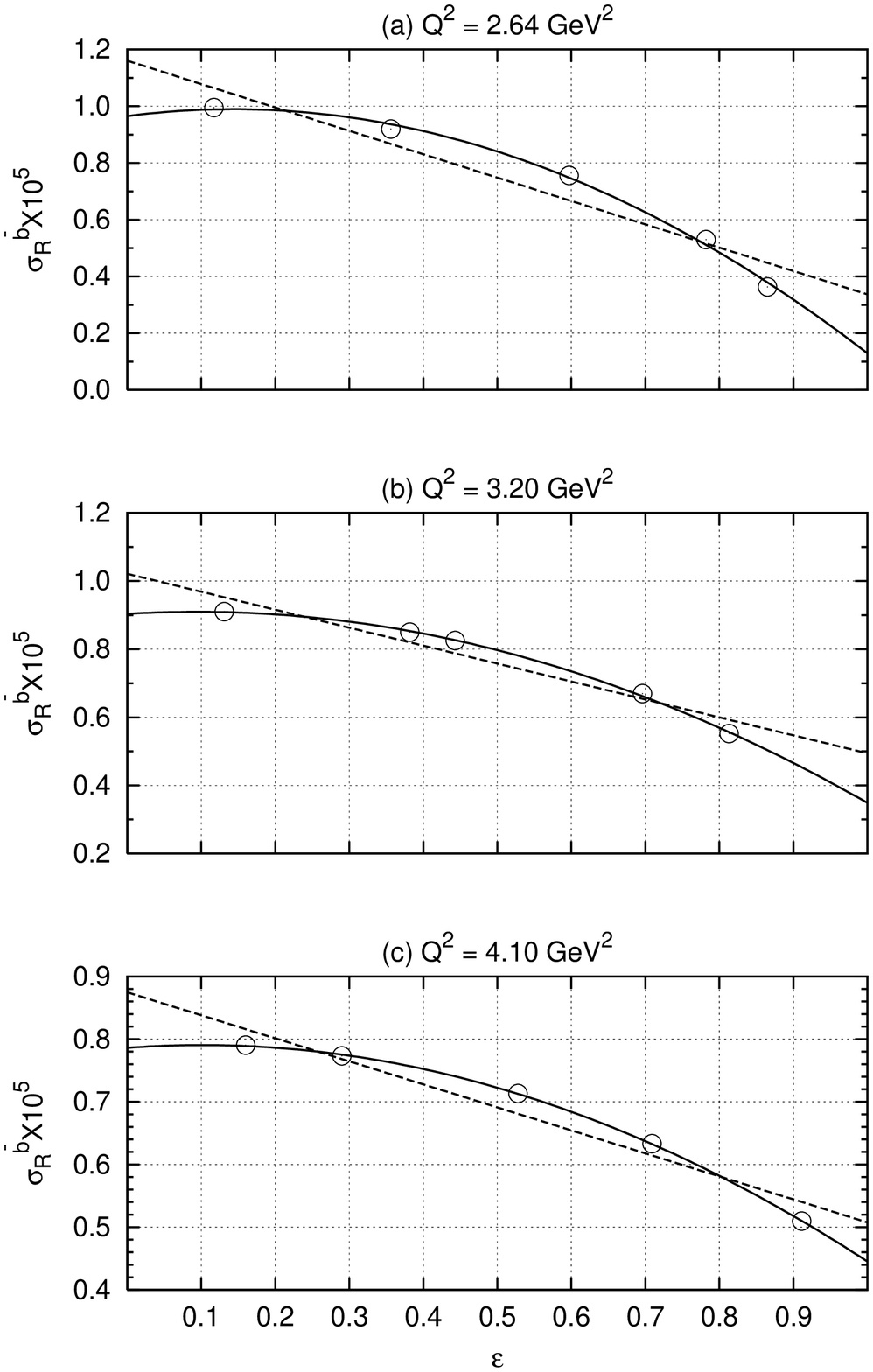}
\caption{\small Contribution of the diagram proportional to $\bar b^2$ to the elastic electron
proton scattering for $Q^2$ = 2.64 GeV$^2$ (a),
3.20 GeV$^2$ (b), 4.10 GeV$^2$ (c). Here $\bar b =1$. The dashed lines represent $f_1$ (eqn. \ref{eq:result.fit.1})
and the solid curves represent $f_2$ (eqn. \ref{eq:result.fit.2}). The fitting parameters are given
in Table \ref{tab:bpmfit}.}
\label{fig:bpm}
\end{figure}

\begin{figure}
\hskip 1.5cm
\includegraphics[scale=0.75,angle=0]{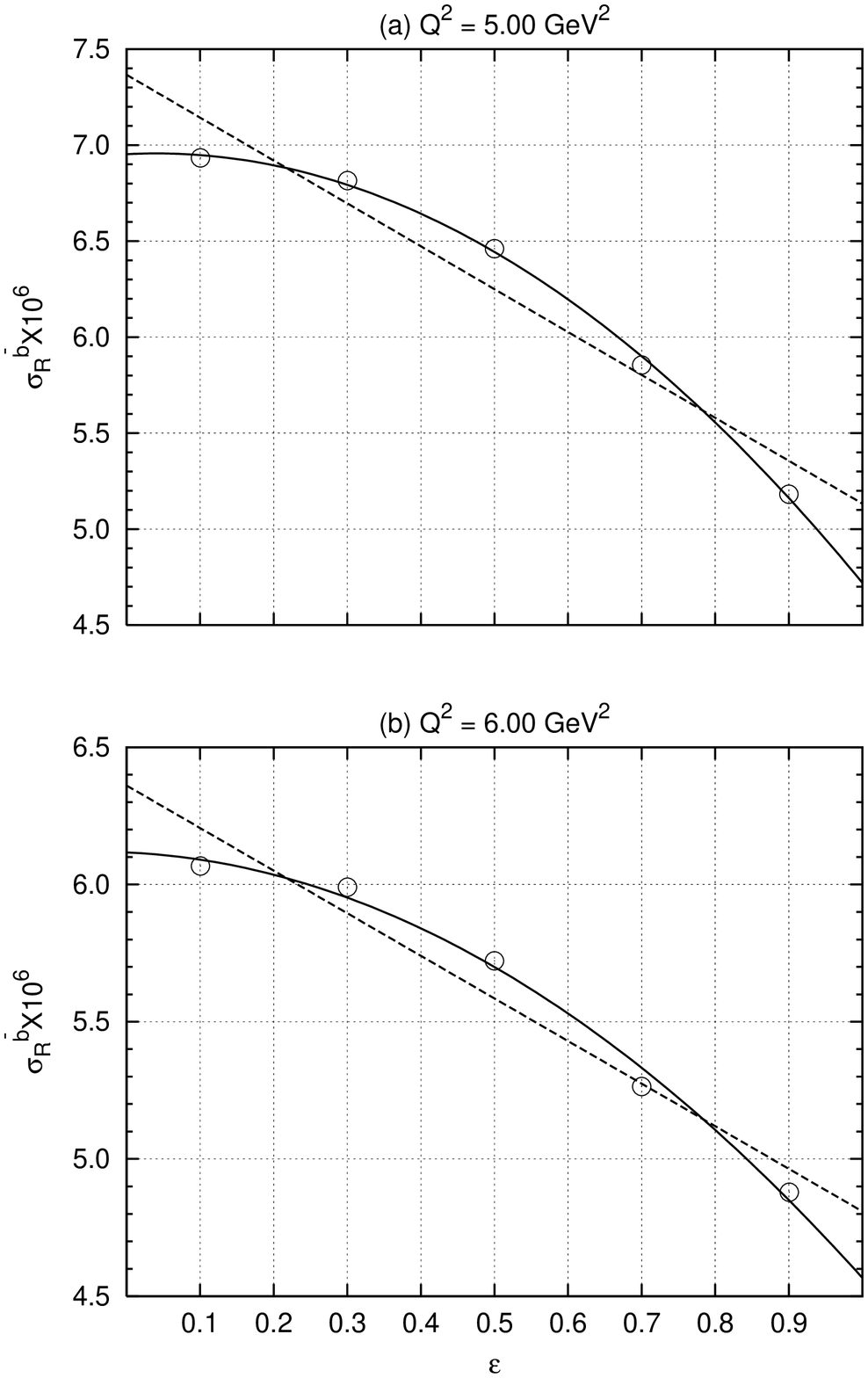}
\caption{\small Contribution of the diagram proportional to $\bar b^2$ to the elastic electron
proton scattering for $Q^2$ = 5 GeV$^2$ (a),
6 GeV$^2$ (b). Here $\bar b =1$. The dashed lines represent $f_1$ (eqn. \ref{eq:result.fit.1})
and the solid curves represent $f_2$ (eqn. \ref{eq:result.fit.2}). The fitting parameters are given
in Table \ref{tab:bpmfit}.}
\label{fig:bpm.2}
\end{figure}

The result of the calculation for the
box and cross-box diagrams is given in Figs. \ref{fig:bcb} and \ref{fig:bcb.2}.
Here we have considered the momentum transfer
$Q^2=2.64$, $3.20$, $4.10$, $5.00$, $6.00$ GeV$^2$. The first three values are same
as those used in the JLAB extraction of form factors using Rosenbluth
separation. The contribution from the diagram proportional to $\bar b^2$ is shown
in Figs. \ref{fig:bpm} and \ref{fig:bpm.2}. Here $\bar b$ is taken as $1$.
We fit $\sigma_R^{BCB}\equiv (\sigma_R^{B}+ \sigma_R^{CB})$ and
$\sigma_R^{\bar b}$ (here and for rest of the section
we use the notations $\sigma_R^{2\gamma}$, $\sigma_R^{BCB}$ to denote the IR removed contributions) to the following functions:
\ba
f_1(\varepsilon) &=& c_1 + c_2\varepsilon\label{eq:result.fit.1}\\
f_2(\varepsilon) &=& d_1 + d_2\varepsilon + d_3\varepsilon^2\label{eq:result.fit.2}
\ea
The values of $c_1$, $c_2$ and $d_1$, $d_2$, $d_3$ for $\sigma_R^{BCB}$
and $\sigma_R^{\bar b}$ are given in Table \ref{tab:bcbfit} and \ref{tab:bpmfit} respectively.
From Table \ref{tab:bcbfit} and \ref{tab:bpmfit} we also see that 
the contribution due to the $\bar b$ term is relatively small as long as the
magnitude of $\bar b$ is of order unity. 
 As the magnitude of $\bar b$ is unknown we shall
assume $\bar b\approx 0$ and take $\sigma_R^{2\gamma} \approx \sigma_R^{BCB}$ for the rest of the section.

\begin{table}
\begin{center}
\begin{tabular}{|c|c|c|c|c|c|}
\hline
$Q^2$&$c_1$&$c_2$&$d_1$&$d_2$&$d_3$\\
\hline
\hline
2.64&$-3.39$&$5.59 $&$-4.06 $&$9.58 $&$-4.04$\\
&$\pm 0.30$&$\pm 0.48$&$\pm 0.12$&$\pm 0.59$&$ \pm 0.58$ \\
\hline
3.20&$-2.31$&$3.99$&$-2.94$&$7.53$&$-3.69$\\
&$\pm 0.25$&$\pm 0.45$&$\pm 0.08$&$\pm 0.38$&$ \pm 0.38$ \\
\hline
4.10&$-1.20$&$2.23$&$-1.61$&$4.30$&$-1.95$\\
&$\pm 0.15$&$\pm 0.25$&$\pm 0.11$&$\pm 0.47$&$ \pm 0.43$ \\
\hline
$5.00$&$-0.79$&$1.63$&$-1.13$&$3.59$&$-1.96$\\
&$\pm 0.16$&$\pm 0.28$&$\pm 0.10$&$\pm 0.47$&$\pm 0.46$\\
\hline
$6.00$&$-0.44$&$1.10$&$-0.69$&$2.55$&$-1.45$\\
&$\pm 0.12$&$\pm 0.21$&$\pm 0.08$&$\pm 0.40$&$\pm 0.39$\\
\hline
\end{tabular}
\end{center}
\caption{\small Values of fitting parameters for ($\sigma_R^B + \sigma_R^{CB}$).
All numbers have been scaled by $10^{4}$. The parameters $c_1$ and $c_2$
are defined at eqn. \ref{eq:result.fit.1} and $d_1$, $d_2$ and $d_3$ are defined
at eqn. \ref{eq:result.fit.2}. The standard errors
obtained in fitting the result are also shown.
Figs. \ref{fig:bcb} and \ref{fig:bcb.2} shows
the functions $f_1$ and $f_2$ for this case. }
\label{tab:bcbfit}
\end{table}

\begin{table}
\begin{center}
\begin{tabular}{|c|c|c|c|c|c|}
\hline
$Q^2$&$c_1$&$c_2$&$d_1$&$d_2$&$d_3$\\
\hline
\hline
2.64&$11.60$&$-8.23$&$9.65$&$3.37$&$-11.73$\\
&$\pm 0.86$&$\pm 1.41$&$\pm 0.37$&$\pm 1.81$&$ \pm 1.79$ \\
\hline
3.20&$10.21$&$-5.27$&$9.04$&$1.29$&$-6.84$\\
&$\pm 0.45$&$\pm 0.83$&$\pm 0.12$&$\pm 0.59$&$ \pm 0.60$ \\
\hline
4.10&$8.75$&$-3.67$&$7.86$&$0.86$&$-4.27$\\
&$\pm 0.31$&$\pm 0.52$&$\pm 0.04$&$\pm 0.16$&$ \pm 0.14$ \\
\hline
$5.00$&$7.37$&$-2.23$&$6.95$&$0.20$&$-2.43$\\
&$\pm 0.19$&$\pm 0.34$&$\pm 0.06$&$\pm 0.29$&$\pm 0.28$\\
\hline
$6.00$&$6.36$&$-1.55$&$6.12$&$-0.12$&$-1.43$\\
&$\pm 0.12$&$\pm 0.21$&$\pm 0.09$&$\pm 0.44$&$\pm 0.42$\\
\hline
\end{tabular}
\end{center}
\caption{\small Values of fitting parameters for $\sigma_R^{\bar b}$ with
$\bar b = 1$.
All numbers have been scaled by $10^{6}$. The standard errors
obtained in fitting the result are also shown. The parameters $c_1$ and $c_2$
are defined at eqn. \ref{eq:result.fit.1} and $d_1$, $d_2$ and $d_3$ are defined
at eqn. \ref{eq:result.fit.2}. Figs. \ref{fig:bpm} and \ref{fig:bpm.2} shows
the functions $f_1$ and $f_2$ for this case. }
\label{tab:bpmfit}
\end{table}

\begin{figure}[h]
\hskip 01cm
\includegraphics[scale=0.55,angle=270]{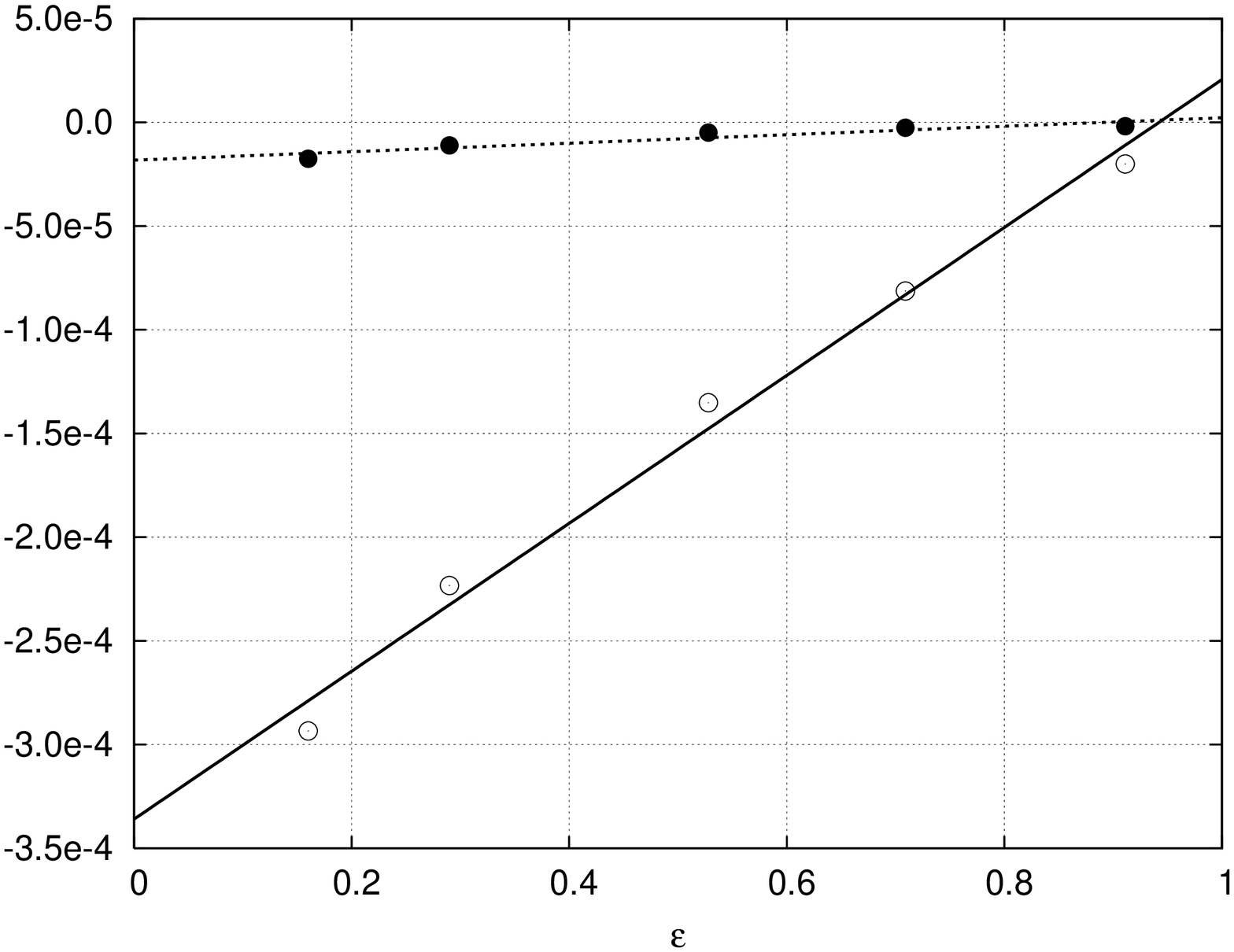}
\caption{\small The contribution to the reduced cross section  
coming from terms proportional to
$F_2(l)F_2(q-l)$ in $\mathcal{M}_{CB}$ and $\mathcal{M}_B$ for $Q^2=4.10$ GeV$^2$,
$\mu^2=0.005$ GeV$^2$ (filled circles). The unfilled circles represent the total contribution
coming from both the form factors $F_1$ and $F_2$ 
for the same $Q^2$ and $\mu^2$.}
\label{fig:f2f2}
\end{figure}

Fig. \ref{fig:f2f2} shows the contribution of the dimension five operator
proportional to $F_2$
to the reduced cross section. This contribution is obtained from the box
and cross-box diagrams. For comparison we also show the
total contribution of both these diagrams. The IR $\mu^2$ dependence is not
removed in this calculation and the parameters chosen are
$Q^2=4.10$ GeV$^2$, $\mu^2=0.005$ GeV$^2$. We find that the contribution
from terms proportional to $F_2 \times F_2$
is much smaller compared to the total contribution, justifying the truncation
of our action to only operators of dimension 5.

To obtain the corrected $\sigma_R$ we subtract the linear fit to
$\sigma_R^{2\gamma}$, $f_1^{2\gamma}(\varepsilon)$
(see eqn. \ref{eq:result.fit.1},
Table \ref{tab:bcbfit}), from a linear fit to $\sigma_R^{LT}(\varepsilon)$ given by:
\be
\sigma_R^{LT}(\varepsilon) = \mathcal{G}_0(1 + \varepsilon\mathcal{G}_1).
\ee
Then the corrected reduced cross section,
\be
\bar{\sigma}_R(\varepsilon) \equiv \sigma_R^{LT}-\sigma_R^{2\gamma} = (\mathcal{G}_0 - c_1^{2\gamma}) + \varepsilon(\mathcal{G}_0\mathcal{G}_1 - c_2^{2\gamma}). \label{eq:sgmcor1}
\ee

\begin{figure}
\hskip 1.5cm
\includegraphics[scale=0.75,angle=0]{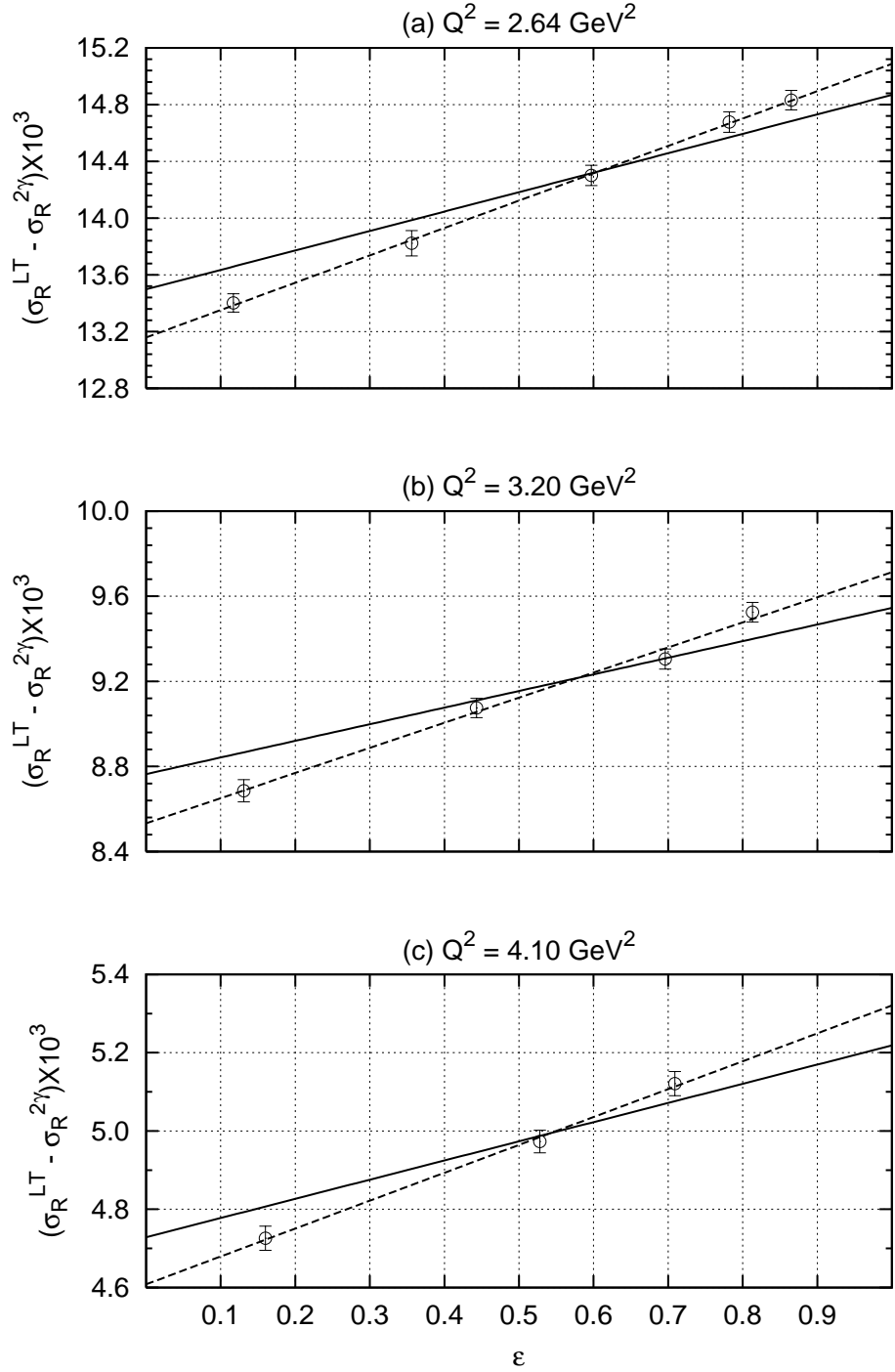}
\caption{\small Corrected cross section, $\bar{\sigma}_R$ (solid lines) obtained using eqn. \ref{eq:sgmcor1}
with $\bar b = 0$ for $Q^2$ = 2.64 GeV$^2$ (a), 3.20 GeV$^2$ (b), 4.10 GeV$^2$ (c). The unfilled circles represent the data
points obtained by Rosenbluth separation method at JLAB and the dashed lines are the straight line fits to these.}
\label{fig:bcbbpm}
\end{figure}

\begin{figure}[!h]
\hskip 1cm
\includegraphics[scale=0.55,angle=270]{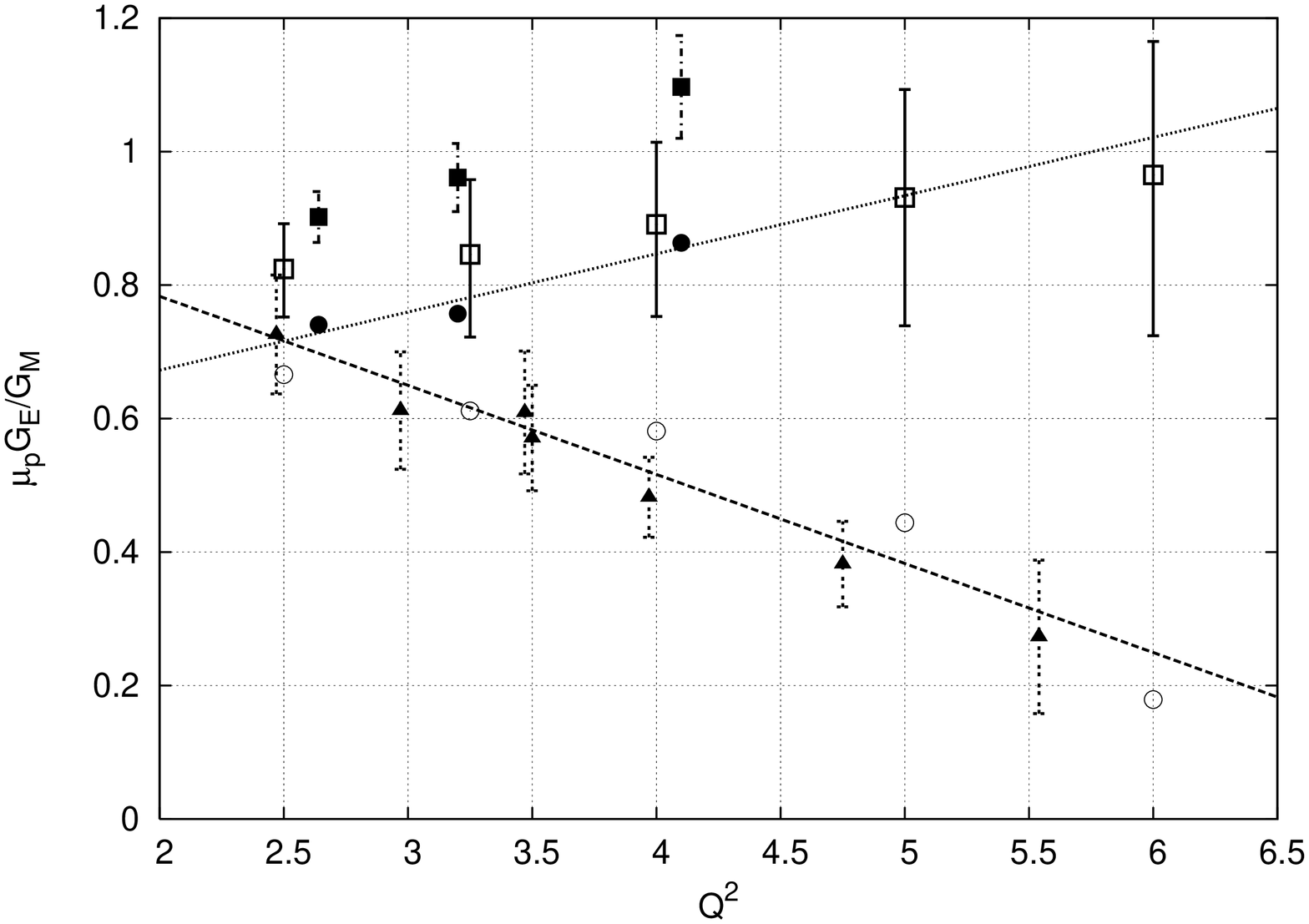}
\caption{\small The ratio, $\mu_pG_E/G_M$ obtained by
polarization transfer technique at JLAB (filled triangles) and
Rosenbluth separation technique at SLAC (unfilled squares) and JLAB (filled
squres). The ratio after correcting for the two photon exchange contribution
is also shown. The filled circles are the corrected JLAB Rosenbluth data and
the dotted line is the best fit through these points. The unfilled circles
are the corrected SLAC Rosenbluth data and
the dashed is line the best fit through these points.
}
\label{fig:mugegmratio1}
\end{figure}
In Fig. \ref{fig:bcbbpm} we plot $\bar{\sigma}_R(\varepsilon)$ for different $Q^2$.
We determine the corrected form factors $\bar{G}_M$ and $\bar{G}_E$ by:
\ba
\bar{G}_M = \frac1\tau\sqrt{\mathcal{G}_0 - c_1^{2\gamma}}\\
\bar{G}_E = \sqrt{\mathcal{G}_0\mathcal{G}_1 - c_2^{2\gamma}}
\ea
Fig. \ref{fig:mugegmratio1}
shows how the ratio $\mu_pG_E/G_M$ is modified by the two photon exchange contributions.
The SLAC Rosenbluth data after applying the two photon exchange correction is
shown by the unfilled circles. The dotted line represents the best linear fit
through this data.
We find that the two photon exchange
correction completely explains the difference between the
SLAC Rosenbluth separation data and the JLAB polarization transfer data.
However it is not able to explain the difference between the JLAB Rosenbluth
and polarization transfer results. The corrected JLAB Rosenbluth data is shown
by filled circles. The JLAB Rosenbluth data lies systematically above the
SLAC data.

\section{Conclusions}

In this paper we have constructed a nonlocal Lagrangian to model
the electromagnetic interaction of proton. The model is invariant
under a nonlocal form of gauge transformations and incorporates
all operators up to dimension five. The model displays the
standard electromagnetic vertex of an on-shell proton. The
dimension five operators also contain an operator with an unknown
coefficient whose value can be extracted experimentally. We use
this model to compute the two photon exchange diagrams
contributing to elastic scattering of electron with proton. The
calculation requires the proton form factors in the entire
kinematic range.
We find that the two photon exchange diagram
contribution to the reduced cross section $\sigma_R$ shows a
slightly non-linear dependence on the longitudinal polarization of the
photon $\varepsilon$. The non-linearity seen is within the experimental error bars
of the Rosenbluth data. We apply the correction due to two photon exchange
contributions to both the SLAC and JLAB Rosenbluth separation data.
The resulting cross section for the SLAC data is completely consistent with
the JLAB polarization transfer results. However the JLAB Rosenbluth data
still shows a large deviation.
It, therefore, appears that the two photon exchange is able to
explain the difference in the experimental extraction of proton
electromagnetic form factor $G_E$ using the Rosenbluth separation
and polarization transfer techniques if we accept the SLAC Rosenbluth
data, which is available over a larger momentum range.

\section{Appendices}
\subsection{Appendix 1}

\medskip
\noindent We shall show in this appendix that the first term in eq.
\ref{eqn:bprime} proportional to
\[
\mathcal{\overline{\psi}}'\left(\sigma_{\mu\nu}f'_{2}\left[\frac{\partial^{2}}{\Lambda^{2}}\right]F^{\mu\nu}\right)^{2}\psi'\]
 does not contribute to the 2-photon matrix element in the 1-loop
approximation in the Feynman gauge in the zero electron mass limit
$m_{e}=0$. For this purpose we write the term
as\[
{1\over 2}\mathcal{\overline{\psi}}'\left(\sigma_{\mu\nu}\sigma_{\lambda\rho}+\sigma_{\lambda\rho}\sigma_{\mu\nu}\right)f'_{2}\left[\frac{\partial^{2}}{\Lambda^{2}}\right]F^{\mu\nu}f'_{2}\left[\frac{\partial^{2}}{\Lambda^{2}}\right]F^{\lambda\rho}\psi'\]
We then note that
\begin{itemize}
\item
$\sigma_{\mu\nu}\sigma_{\lambda\rho}+\sigma_{\lambda\rho}\sigma_{\mu\nu}=$ a linear combination of $\mathbb{I}$ and $\gamma_5
 =  \alpha\left(g_{\mu\lambda}g_{\nu\rho}-g_{\mu\rho}g_{\nu\lambda}\right)\mathbb{I}+i\beta\xi_{\mu\nu\lambda\rho}\gamma_5$, where
$\alpha,\beta$ are constants and, in particular, \emph{no}
$\sigma-$terms appear.
\item The Feynman integral has no dependence on both $p$ and $p'$ .
\item Thus, the result for the 2-photon exchange diagram is of the form:\[
\overline{u}(k')\gamma_{\mu}\gamma_{\alpha}\gamma_{\nu}u(k)\overline{U}(p')[\mathbb{I},\gamma_5]U\left(p\right)\times
I^{\mu\alpha\nu}(k,k')\]
On simplification, this becomes \[
\overline{u}(k')\left\{
g_{\mu\alpha}\gamma_{\nu}+g_{\nu\alpha}\gamma_{\mu}-g_{\mu\nu}\gamma_{\alpha}+4i\xi_{\mu\alpha\nu\beta}\gamma^{\beta}\gamma_5\right\}
u(k)\overline{U}(p')[\mathbb{I},\gamma_5]U\left(p\right)\times
I^{\mu\alpha\nu}(k,k')\] Now,
$g_{\mu\alpha}I^{\mu\alpha\nu}(k,k')$ is a linear combination of
terms that are $\sim k^{\nu}\, or\,\, k'^{\nu}$. Both of these terms give zero.
Similar logic applies to all other terms.
\end{itemize}

\subsection{Appendix 2: Model for the Form Factors}

\noindent
The fits for $G_M/\mu_p$ and $G_E$ are given by:
\ba
\frac{G_M(q^2)}{\mu_p} = \sum_{a=1}^4\frac{A^{'}_a}{(q^2-m_a^2+im_a\Gamma^{'}_a)}\label{gmbymu}\\
G_E(q^2) = \sum_{a=1}^6\frac{B^{'}_a}{(q^2-m_a^2+im_a\Gamma^{'}_a)}\label{ge}.
\ea
We have considered two fits for $G_E$. The values of the masses and the parameters are tabulated in Table \ref{tab1} (Model I)
and Table \ref{tab2} (Model II).

\begin{table}[!h]
\hskip 4cm
\begin{tabular}{|c|c|c|c|c|}
\hline
$a$&$A^{'}_a$&$B^{'}_a$&$m_a$&$\Gamma^{'}_a$\\
\hline
\hline
1&$-2.882564$&$-3.177877$&0.8084&0.2226\\
&$+i\, 1.944314$&$+i\,2.123389$&&\\
\hline
2&$2.882564 $&$3.177877$&0.9116&0.1974\\
&$- i\, 1.944314$&$ - i\,2.123389$&&\\
\hline
3&$-1.064011$&$-0.608148$&1.274&0.5712\\
&$ - i\,3.216318$&$ - i\,5.685885$&&\\
\hline
4&$1.064011$&$0.608148$&1.326&0.5488\\
&$+ i\,3.216318$&$ + i\,5.685885$&&\\
\hline
5&0&$3.211388$&1.96&1.02\\
&&$+ i\,0.693412$&&\\
\hline
6&0&$- i\,0.693412$&2.04&0.98\\
&&$ - i\,0.693412$&&\\
\hline
\end{tabular}
\caption{\small Masses, widths and parameter values for $G_M/\mu_p$ and $G_E$ fits (Model I).
$A^{'}$s and $B^{'}$s are defined in eq. \ref{gmbymu} and \ref{ge}.}
\label{tab1}
\end{table}

\begin{table}[!h]
\hskip 4cm
\begin{tabular}{|c|c|c|c|c|}
\hline
$a$&$A^{'}_a$&$B^{'}_a$&$m_a$&$\Gamma^{'}_a$\\
\hline
\hline
1&$-2.882564$&$-3.392256$&0.8084&0.2226\\
&$+ i\, 1.944314$&$+i\,2.194129$&&\\
\hline
2&$2.882564$&$3.392256$&0.9116&0.1974\\
&$- i\, 1.944314$&$-i\,2.194129$&&\\
\hline
3&$-1.064011$&$1.224037$&1.274&0.5712\\
&$- i\,3.216318$&$-i\,6.877523$&&\\
\hline
4&$1.064011$&$-1.224037$&1.326&0.5488\\
&$+ i\,3.216318$&$+i\,6.877523$&&\\
\hline
5&0&$1.645805$&2.107&0.663\\
&&$+i\,1.824298$&&\\
\hline
6&0&$-1.645805$&2.193&0.637\\
&&$-i\,1.824298$&&\\
\hline
\end{tabular}
\caption{\small Masses, widths and parameter values for $G_M/\mu_p$
and $G_E$ fits (Model II). $A^{'}$s and $B^{'}$s are defined in eq. \ref{gmbymu} and \ref{ge}.}
\label{tab2}
\end{table}

Using the models for the magnetic and electric form factors we can determine
the Dirac and Pauli form factors.
Let the fits to the form factors $G_M$ and $G_E$ be:
\ba
G_M(q^2) = \sum_{a=1}^6 A_a g_a(q^2)\\
G_E(q^2) = \sum_{a=1}^6 B_a g_a(q^2)
\ea
with $A_a = \mu_pA^{'}_a$ and $B_a = B^{'}_a$. The $g_a$'s are defined by,
\be
g_a(q^2) = \frac{1}{q^2-m_a^2 + i \Gamma_a}.
\ee
The form factors $F_1$ and $F_2$ are given by,
\ba
\kappa_p F_2 = \frac{G_M-G_E}{1+\tau} = 4M_p^2\frac{G_E-G_M}{q^2-4M_p^2} = \sum_a4M_p^2D_a\tilde{g}_a\\
F_1 = G_M-\kappa_p F_2 = \sum_{a=1}^6(C_a g_a - 4M_p^2D_a\tilde{g}_a).
\ea
where $C_a =A_a$, $D_a = B_a - A_a$ and $\tilde{g}_a = g_a/(q^2-4M_p^2)$.
These definitions are convenient in evaluating the two photon exchange
amplitudes. We also have,
\ba
\frac{\kappa_p F_2}{q^2-\mu^2+i\xi}
&=& \sum_{a=1}^6 \frac{4M_p^2D_a}{(q^2-\mu^2+i\xi)(q^2-4M_p^2)(q^2-m_a^2 + i \Gamma_a)} \\
&=& \sum_{i=1}^8 \frac{4M_p^2D'_i}{q^2-m_i^2+i\Gamma_i},
\label{eq:F2_1}
\ea
with $m_7=\mu$, $m_8=2M_p$, $\Gamma_{7}=\xi$ and $\Gamma_{8}=0$. Here
($a=1,2 ...,6$)
\bas
D'_a &=& \frac{D_a}{(4M_p^2-\mu^2+i\xi)}\left[\frac{1}{m_a^2-4M_p^2-i\Gamma_a}-\frac{1}{m_a^2-\mu^2+i\xi-i\Gamma_a} \right],\\
D'_7 &=& \sum_{a=1}^6 \frac{D_a}{(4M_p^2-\mu^2+i\xi)(m_a^2-\mu^2+i\xi-i\Gamma_a)},\\
D'_8 &=& -\sum_{a=1}^6 \frac{D_a}{(4M_p^2-\mu^2+i\xi)(m_a^2-4M_p^2-i\Gamma_a)} = 0.
\eas
The coefficient $D_8'$ is found to be zero and hence the summation in eq.
\ref{eq:F2_1} terminates at $i=7$.
Similarly,
\be
\frac{F_1}{q^2-\mu^2+i\xi}= \sum_{i=1}^8 \frac{C'_i-4M_p^2D'_i}{q^2-m_i^2+i\Gamma_i},
\ee
where
\bas
C'_a &=& \frac{C_a}{m_a^2-\mu^2+i\xi-i\Gamma_a},\\
C'_7 &=& -\sum_{a=1}^6 \frac{C_a}{m_a^2-\mu^2+i\xi-i\Gamma_a},\\
C'_8 &=& 0.
\eas
We can also write $F_1$ and $\kappa_p F_2$ using this general notation.
We find
\ba
\kappa_p F_2 &=& \sum_{i=1}^7 \frac{4M_p^2D''_i}{q^2-m_i^2+i\Gamma_i},\\
F_1 &=& \sum_{i=1}^7 \frac{C''-4M_p^2D''_i}{q^2-m_i^2+i\Gamma_i},
\ea
with
\bas
D''_a &=& \frac{D_a}{m_a^2-4M_p^2+i\xi-i\Gamma_a},\\
D''_7 &=& 0,\\
C''_a &=& A_a,\\
C''_{7} &=& 0.
\eas

\subsection{Appendix 3: Sample Calculation: Box Diagram}

\noindent
Here we present a sample calculation of one of the terms in the Box diagram.
The contribution of the box diagram amplitude to the two photon exchange
cross section is proportional to,
\be
\overline{\mathcal{M}^*_0\mathcal{M}'_B} = i\frac{e^6}{q^2}\sum_{i,j}\mathcal{I}_{B}^{ij}
\ee
where,
\bas
\mathcal{I}_{B}^{ij} &=& \int\frac{d^4l}{(2\pi)^4} \frac{\mathcal{N}^{ij}(l)}{((k-l)^2-m_e^2+i\xi)
((p+l)^2-M_p^2+i\xi)(\tilde{q}^2-m_i^2+i\Gamma_i)}\\
&&\times\frac{1}{(l^2-m_j^2+i\Gamma_j)}
\eas
We can now evaluate this integral by the standard Feynman parametrization
technique. We define
\be
\mathcal{D} =
l^2 + 2l.(x_2p -x_1k-x_3q) + x_3(q^2-m_i^2) - x_4m_j^2+i\xi+i(x_3\Gamma '_i+x_4\Gamma '_j)
\ee
with $\Gamma '_{i} = \Gamma_{i}-\xi$.
We now define the shifted momentum,
$r = l + (x_2p -x_1k-x_3q)$ which gives,
$\mathcal{D} = r^2 - \Delta'$,
with
\bas
\Delta' &=& x_1^2m_e^2+x_2^2M_p^2 -x_3(1-x_1-x_2-x_3)q^2-2x_1x_2EM_p + x_3m_i^2+x_4m_j^2 \\
&-& i\xi - i(x_3\Gamma '_i+x_4\Gamma '_j)
\eas
With this momentum shift the numerator becomes:
\bas
\mathcal{N}^{ij}(l) = \mathcal{N}_0 + r_\mu\mathcal{N}_1^\mu + r_\mu r_\nu\mathcal{N}_2^{\mu\nu}
+ r_\mu r_\nu r_\rho\mathcal{N}_3^{\mu\nu\rho}.
\eas
Hence,
\bas
\mathcal{I}_{B}^{ij} = 6\int_0^1\Pi_{\alpha=1}^4dx_\alpha\delta(\sum_{\alpha=1}^4x_\alpha-1)
\int\frac{d^4l}{(2\pi)^4}\frac{\mathcal{N}_0 + r_\mu\mathcal{N}_1^\mu + r_\mu r_\nu\mathcal{N}_2^{\mu\nu}
+ r_\mu r_\nu r_\rho\mathcal{N}_3^{\mu\nu\rho}}{(r^2-\Delta')^4}.
\eas
As the denominator depends only on the magnitude of $r$,
\bas
\int\frac{d^4r}{(2\pi)^4}\frac{r_\mu\mathcal{N}_1^\mu}{\mathcal{D}^4} &=& 0,\\
\int\frac{d^4r}{(2\pi)^4}\frac{r_\mu r_\nu\mathcal{N}_2^{\mu\nu}}{\mathcal{D}^4} &=&
\int\frac{d^4r}{(2\pi)^4}\frac{\frac14g_{\mu\nu}\mathcal{N}_2^{\mu\nu}r^2}{\mathcal{D}^4},\\
\int\frac{d^4r}{(2\pi)^4}\frac{r_\mu r_\nu r_\rho\mathcal{N}_3^{\mu\nu\rho}}{\mathcal{D}^4} &=& 0.
\eas
Let $\mathcal{N}_2$ be the shorthand notation for $\frac14g_{\mu\nu}\mathcal{N}_2^{\mu\nu}$.
\bas
\therefore \mathcal{I}_{B}^{ij} &=& 6\int_0^1\Pi_{\alpha=1}^4dx_\alpha\delta(\sum_{\alpha=1}^4x_\alpha-1)
\int\frac{d^4r}{(2\pi)^4}\frac{\mathcal{N}_0 + r^2\mathcal{N}_2}{(r^2-\Delta')^4}\\
&=&\frac{i}{16\pi^2}(\mathcal{I}_0-2\mathcal{I}_2)\\
\mbox{with } \mathcal{I}_0 &=& \int_0^1dx_3\int_0^{1-x_3}dx_2\int_0^{1-x_3-x_2}dx_1 \frac{\mathcal{N}_0}{\Delta^2}\\
\mbox{and }  \mathcal{I}_2 &=& \int_0^1dx_3\int_0^{1-x_3}dx_2\int_0^{1-x_3-x_2}dx_1 \frac{\mathcal{N}_2}{\Delta}.
\eas
Here
\bas
\Delta &=& x_1^2m_e^2+x_2^2M_p^2 -x_3(1-x_1-x_2-x_3)q^2-2x_1x_2EM_p + x_3m_i^2\\
&&+(1-x_1-x_2-x_3)m_j^2 - i\xi - i(x_3\Gamma '_i+(1-x_1-x_2-x_3)\Gamma '_j).
\eas
If we neglect the mass of electron then,
\bas
\Delta&\approx& x_2^2M_p^2 -x_3(1-x_1-x_2-x_3)q^2-2x_1x_2EM_p + x_3m_i^2\\
&&+(1-x_1-x_2-x_3)m_j^2 - i\xi - i(x_3\Gamma '_i+(1-x_1-x_2-x_3)\Gamma '_j)\\
&\equiv& Xx_1 + Y,
\eas
where,
\bas
X &=& - 2x_2EM_p + x_3q^2 - m_j^2 + i\Gamma'_j\\
Y &=& x_2^2M_p^2 - x_3(1-x_2-x_3)q^2 + x_3(m_i^2-m_j^2) + (1-x_2)m_j^2 \\
&-&i\xi - ix_3(\Gamma'_i-\Gamma'_j)-i(1-x_2)\Gamma'_j\,.
\eas
 $\mathcal{N}_1,\,\mathcal{N}_2$ can be written as:
\bas
\mathcal{N}_0 &=& Z_3+Z_4x_1+Z_5x_1^2\\
\mathcal{N}_2 &=& Z_1+Z_2x_1.
\eas
Then
\bas
\mathcal{I}_0 &=& \int_0^1dx_3\int_0^{1-x_3}dx_2\int_0^{1-x_3-x_2}dx_1 \frac{Z_3+Z_4x_1+Z_5x_1^2}{(Xx_1+Y)^2}\\
\mbox{and }\mathcal{I}_2 &=& \int_0^1dx_3\int_0^{1-x_3}dx_2\int_0^{1-x_3-x_2}dx_1 \frac{Z_1+Z_2x_1}{(Xx_1+Y)}.
\eas
The $x_1$ integration can be done analytically to obtain:
\ba
\nn \mathcal{I}^{ij}_B &=& \frac{i}{16\pi^2}\int_0^1dx_3\int_0^{1-x_3}dx_2\left[\frac{-2Z_1X^2 + X(2Z_2Y + Z_4)-2Z_5Y}{X^3}\,\ln\left(\frac{XL+Y}{Y}\right)\right.\\
\nn &&\left.+\frac{L}{X^2Y(XL+Y)}\left\{2Z_5Y^2 + X^2(Z_3-2Z_2LY) -XY(2Z_2Y+Z_4-Z_5L)\right\} \right],\\
\ea
where $L = 1 - x_3 - x_2$. $Z_i$'s are obtained using FORM \cite{FORM} and $I^{ij}_B$'s are numerically computed using Gauss-Legendre integration technique
\cite{NR}.


\end{document}